\documentclass[twocolumn]{aastex631}

\newcommand{\kB}{k_{\rm B}}
\newcommand{\mproton}{m_{\rm p}}
\newcommand{\Tvir}{T_{\rm vir}}
\newcommand{\Vvir}{V_{\rm vir}}
\newcommand{\fb}{f_{\rm b}}
\newcommand{\Rvir}{R_{\rm vir}}

\shorttitle{Circumgalactic turbulence, atomic cooling and galaxy formation}
\shortauthors{Pandya et al.}

\graphicspath{{./}{figures/}}

\begin{document}

\title{A unified model for the co-evolution of galaxies and their circumgalactic medium: the relative roles of turbulence and atomic cooling physics}

\correspondingauthor{Viraj Pandya}
\email{vgp2108@columbia.edu}

\author{Viraj Pandya}
\altaffiliation{Hubble Fellow}
\affiliation{Columbia Astrophysics Laboratory, Columbia University, 550 West 120th Street, New York, NY 10027, USA}
\affiliation{Center for Computational Astrophysics, Flatiron Institute, New York, NY 10010, USA}

\author{Drummond B. Fielding}
\affiliation{Center for Computational Astrophysics, Flatiron Institute, New York, NY 10010, USA}

\author{Greg L. Bryan}
\affiliation{Department of Astronomy, Columbia University, 550 West 120th Street, New York, NY 10027, USA}
\affiliation{Center for Computational Astrophysics, Flatiron Institute, New York, NY 10010, USA}

\author{Christopher Carr}
\affiliation{Department of Astronomy, Columbia University, 550 West 120th Street, New York, NY 10027, USA}

\author{Rachel S. Somerville}
\affiliation{Center for Computational Astrophysics, Flatiron Institute, New York, NY 10010, USA}

\author{Jonathan Stern}
\affiliation{School of Physics, Astronomy, Tel Aviv University, Tel Aviv 69978, Israel}

\author{Claude-Andr\'e Faucher-Gigu\`ere}
\affiliation{Department of Physics and Astronomy and CIERA, Northwestern University, 1800 Sherman Ave, Evanston, IL 60201, USA}

\author{Zachary Hafen}
\affiliation{Department of Physics and Astronomy, University of California Irvine, Irvine, CA 92697, USA}

\author{Daniel Angl\'es-Alc\'azar }
\affiliation{Department of Physics, University of Connecticut, 196 Auditorium Road, U-3046, Storrs, CT 06269-3046, USA}
\affiliation{Center for Computational Astrophysics, Flatiron Institute, New York, NY 10010, USA}

\author{John C. Forbes}
\affiliation{Center for Computational Astrophysics, Flatiron Institute, New York, NY 10010, USA}

\begin{abstract}

The circumgalactic medium (CGM) plays a pivotal role in regulating gas flows around galaxies and thus shapes their evolution. However, the details of how galaxies and their CGM co-evolve remain poorly understood. We present a new time-dependent two-zone model that self-consistently tracks not just mass and metal flows between galaxies and their CGM but also the evolution of the global thermal and turbulent kinetic energy of the CGM. Our model accounts for heating and turbulence driven by both supernova winds and cosmic accretion as well as radiative cooling, turbulence dissipation, and halo outflows due to CGM overpressurization. We demonstrate that, depending on parameters, the CGM can undergo a phase transition (``thermalization'') from a cool, turbulence-supported phase to a virial-temperature, thermally-supported phase. This CGM phase transition is largely determined by the ability of radiative cooling to balance heating from supernova winds and turbulence dissipation. We perform an initial calibration of our model to the FIRE-2 cosmological hydrodynamical simulations and show that it can approximately reproduce the baryon cycles of the simulated halos. In particular, we find that, for these parameters, the phase transition occurs at high-redshift in ultrafaint progenitors and at low redshift in classical $M_{\rm vir}\sim10^{11}M_{\odot}$ dwarfs, while Milky Way-mass halos undergo the transition at $z\approx0.5$. We see a similar transition in the simulations though it is more gradual, likely reflecting radial dependence and multi-phase gas not captured by our model. We discuss these and other limitations of the model and possible future extensions.

\end{abstract}


\section{Introduction}
Galaxy formation is the result of numerous physical processes spanning orders of magnitude in both spatial and temporal scales. These include the gravitational collapse of dark matter halos within the large-scale cosmic web, the accretion of gas into those halos, its radiative cooling and inflow towards the center of the halo potential well, the subsequent formation of stars, and the deposition of mass, momentum, energy and metals back into the system via feedback from stellar and black hole evolution. All of these physical processes (and others) leave an imprint on the diffuse volume-filling gas surrounding galaxies within halos known as the circumgalactic medium \citep[CGM; for a recent review, see][]{tumlinson17}. While the CGM regulates the large-scale flows of gas in and out of halos and therefore plays a crucial role in shaping the evolution of galaxies, the small-scale processes occurring within galaxies like star formation and supernova-driven winds can themselves dramatically influence the large-scale physical conditions of the CGM. Thus the properties of galaxies and their CGM must be intimately connected.

Observational efforts to probe the physical conditions of the CGM around galaxies of different types have uncovered many tantalizing trends. Around our own Milky Way (MW) Galaxy, a combination of X-ray and UV studies in both absorption and emission have detected a substantial reservoir of cool ($T\sim10^4-10^5$ K), warm ($T\sim10^5-10^6$ K) and hot ($T\gtrsim10^6$ K) gas which together with certain assumptions may fully account for the missing baryons \citep[e.g.,][]{sembach03,bregman07,anderson10,henley10,gupta12,fang15,das21}. There is also a substantial population of cold ($T\lesssim10^4$ K) ``high-velocity'' clouds detected via their 21 cm emission around the MW, but their total mass is a small fraction of the baryon budget \citep[see review by][and references therein]{putman12}. Around other nearby galaxies, constraints mainly come from UV quasar absorption line studies, which reveal that cool and warm gas may be ubiquitous in the CGM of local galaxies spanning a range of stellar masses \citep[e.g.,][]{prochaska11,tumlinson11,werk14,bordoloi14,stern16,werk16}. However, uncertainties in detailed ionization modeling, the abundances of individual elements, the assumed extent of the CGM, variations in the physical conditions along different lines of sight for a given CGM, the limited number of such sightlines, and the scarcity of constraints on the possibly dominant hot phase from X-rays \citep[but see][]{strickland04,tullmann06,anderson11,bogdan13} significantly hamper our ability to draw strong conclusions about the nature of the CGM. At higher redshifts, there are a wealth of constraints from both absorption line studies \citep[e.g.,][]{steidel10,bordoloi11,rudie12,rudie19,burchett19,chen20} and CGM emission maps \citep[e.g.,][]{leclercq17,wisotzki16,wisotzki18,leclercq20}. Even more CGM constraints are expected in the future from observations of the Sunyaev-Zel'dovich effect around galaxy-scale halos \citep{mroczkowski19}, localized fast radio bursts \citep{prochaska19,wu22}, and the next-generation of ground- and space-based observatories. 

In order to interpret all of these data and understand how galaxies and their CGM co-evolve, we must turn to theoretical models. These can roughly be grouped into three categories: (1) hydrodynamical simulations, (2) 1D models that describe the properties of the CGM at a single moment in time, and (3) simplified time-dependent multi-zone models that model the co-evolution of both galaxies and their CGM. Of these, hydrodynamical simulations are perhaps the most appealing because they attempt to self-consistently track the thermodynamics of gas flows in and around galaxies with fewer assumptions than the other two approaches \citep[see the recent review by][]{naab17}. Of course, they still suffer from uncertainties due to their implementation of unresolved ``subgrid physics'' such as star formation, turbulence, metal mixing, etc., which arise because of limitations in both resolution and physical understanding. In addition, the complexity and cost of these simulations demands the development of simpler 1D and multi-zone models to distill their key predictions. Nevertheless, both idealized and cosmological simulations are useful for testing our understanding of the physical principles that might govern the CGM--galaxy connection. Idealized simulations may focus on small patches of the CGM to understand the microphysics of turbulence and multiphase gas \citep[e.g.,][]{mccourt12,mccourt18,fielding20,abruzzo22,gronke22} but can also model global scales to understand the CGM \citep[e.g.,][]{sharma12,fielding17a,stern19,stern20,lochhaas20,li20b}. On the other hand, fully cosmological simulations can provide insights into the nature of cosmic accretion \citep[e.g.,][see also the time-dependent 1D simulations by \citealt{birnboim03,dekel06}]{keres05,dekel09,vandevoort11,fauchergiguere11,nelson13,anglesalcazar17,hafen20,forbes23}, the contribution of satellites to the cool CGM reservoir \citep[e.g.,][]{fauchergiguere16,hafen19,fielding20b}, and how the conditions of the CGM affect the formation of galactic structure \citep[e.g.,][]{stern21,gurvich22,hafen22}. An increasingly popular approach is also to forward model CGM observables using cosmological simulations \citep[e.g.,][]{vandevoort13,corlies16,oppenheimer18,lokhorst19,defelippis21,moser22}.

On the instantaneous 1D modeling side, the three main physical principles that are usually implemented to describe the CGM in both simulations and observations are hydrostatic equilibrium models \citep[HSE; e.g.,][]{faerman17,qu18,faerman20}, steady-state cooling flow solutions \citep[e.g.,][respectively, for applications to group/cluster- and galaxy-scale halos]{fabian94,stern19}, and precipitation models \citep[e.g.,][]{murray90,maller04,mccourt12,sharma12,voit15}. HSE models imagine that radiative losses in the CGM are balanced by energy input from feedback as well as non-thermal sources of pressure support such as turbulence, cosmic rays and magnetic fields. Challenges remain in extending these ``quasi-hydrostatic'' models to different halo mass scales where non-equilibrium processes may be important. Precipitation models are a subset of HSE models which assume that cool gas condenses out of a predominantly hot background CGM whenever and wherever the ratio of the gas cooling time to freefall time drops below $\sim10$, at which point thermal instabilities can develop. Steady-state cooling flow solutions are compelling in that they solve the fluid equations for a spherically symmetric distribution of gas experiencing gravitational collapse due to the loss of entropy from cooling. This approach can predict the density and temperature profiles of the CGM assuming that feedback effects are negligible. None of these idealized 1D models simultaneously model the galaxy formation process and self-consistently predict the different mass, metal and energy source and sink terms for the CGM as a function of time.

In this paper we focus on the third category of simplified but time-dependent multi-zone models. Unlike 1D models that are continuous in the length dimension or 0D models that only consider a single zone, multi-zone models use a system of coupled ordinary differential equations (ODEs) to predict the state of multiple discrete components of a physical system owing to the flow of matter and energy between them. These models allow one to predict the buildup of mass and metals in the interstellar medium (ISM), CGM and long-lived stars for large populations of halos in a cosmological context with a high degree of computational efficiency. There are simplified approaches called ``bathtub'' models that include limited treatment of the detailed underlying physics and are usually restricted to one zone \citep[the ISM;][]{erb08,bouche10,dave12,lilly13,forbes14b,dekel14,rodriguezpuebla16,tacconi20,kravtsov22} as well as more comprehensive semi-analytic models (SAMs) that implement a wider range of physical processes over three zones \citep[ISM, CGM and intergalactic medium; e.g., see early papers by][and the reviews by \citealt{benson10} and \citealt{somervilledave15}]{whiterees78,whitefrenk91,kauffmann93,somerville99,cole00}. While these existing approaches have shown great success in being able to predict the properties of galaxies at a range of redshifts, their underlying CGM framework usually traces back to \citet{whitefrenk91} who assume that the thermodynamics of the CGM is coupled to that of the dark matter in the sense that the CGM temperature everywhere must be the same as the halo virial temperature. There have been a few efforts to develop an updated CGM basis for SAMs but these appear to not have become the norm \citep[e.g.,][]{lu11,benson11,cousin15,hou18}. With the ever-increasing complexity of high-resolution hydrodynamical simulations and the growing abundance of observational constraints on the CGM, it is high time to revisit the foundation of SAMs which ultimately lies in the assumed CGM model since that regulates the gas flow cycle into and out of galaxies and halos. 

Here we will present a new time-dependent two-zone model that tracks not only mass and metal flows between galaxies and their CGM but also energy flows. Our model assumes that both SN-driven winds and cosmic accretion deposit thermal energy and drive turbulence in the CGM. This then lets us self-consistently predict the global average temperature and characteristic turbulent velocity of the CGM. Thus we will show how the thermodynamics of the CGM can be decoupled from that of the underlying dark matter and what the implications are for the phase of the CGM as a function of cosmic time. In particular, we will elucidate the relative roles of turbulence and atomic cooling physics in regulating the evolution of the CGM and hence also galaxy formation. We will use the high-resolution cosmological hydrodynamical ``zoom-in'' simulations from the FIRE-2 suite \citep{hopkins18} to calibrate our model, although we note that the model could be calibrated to other simulations, or even, given enough data, to observations. In a companion paper \citep{carr23}, we use the purely thermal limit of this kind of model to predict the stellar-to-halo-mass relation and ISM gas fractions that can be compared to empirically-derived constraints from observations, finding that lower mass galaxies require winds that carry a larger fraction of their supernovae energy. After demonstrating the power of our new approach, we will discuss several ways in which the model can be extended in the future to summarize the essential physics of galaxy formation and interpret a wide variety of observational data on both galaxies and their CGM.

This paper is organized as follows. In Section 2, we define the state variables, ODEs and assumptions of our model. In Section 3, we apply the model to an idealized $z=0$ MW-mass CGM to explore its equilibrium behavior and effect of parameter variations. In Section 4, we describe how we measure various galaxy and CGM properties from the FIRE-2 simulations for model calibration and validation purposes. In Section 5, we compare the predictions of the model to the FIRE-2 simulations in terms of the mass assembly histories, baryon cycles and CGM energetics for individual halos as a function of time. After a discussion in Section 6, we conclude in Section 7. We assume a standard \citet{planck15} cosmology with $h=0.6774$, $\Omega_{\rm m,0}=0.3075$, $\Omega_{\rm \Lambda,0}=0.691$ and $\Omega_{\rm b,0}=0.0486$.

\section{Model Description}
\begin{figure*}
\centering
\includegraphics[width=\hsize]{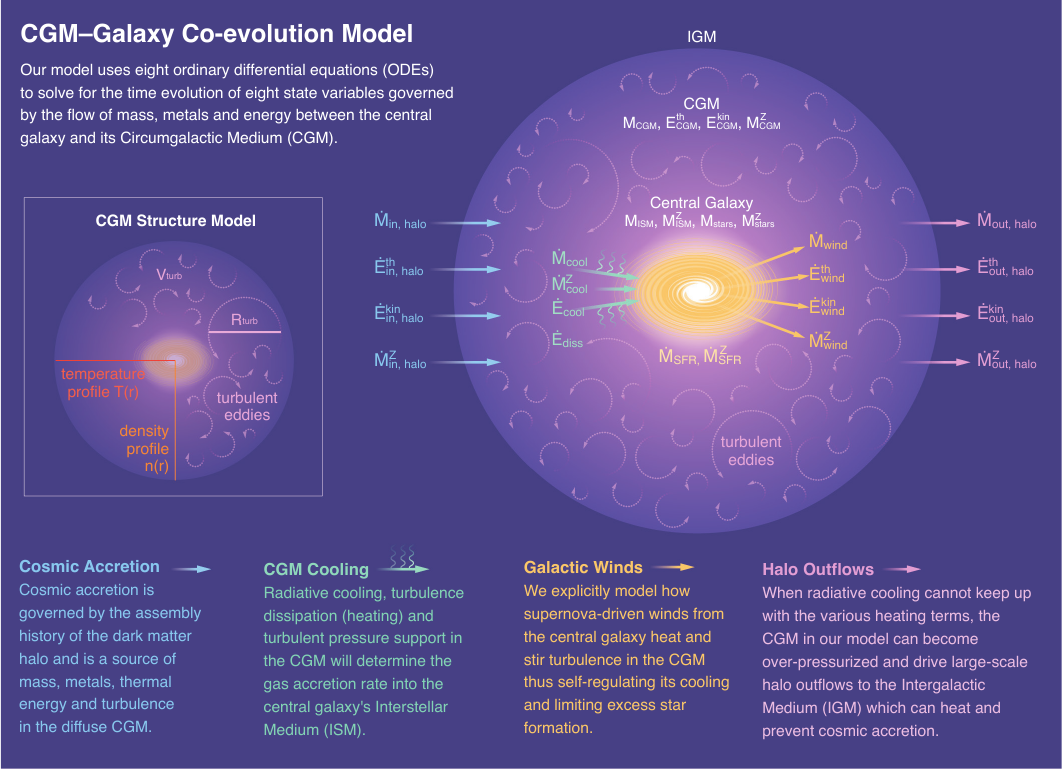}
\caption{An illustration of our new CGM--galaxy co-evolution model. The CGM is described by four state variables (the total CGM mass, thermal energy, turbulent kinetic energy and metal mass) and the galaxy is described by four additional state variables (the masses and metal masses of the ISM and long-lived stellar population). Each of these state variables is evolved according to a system of coupled ODEs as defined in Equation \ref{eqn:system} and as illustrated with the flux arrows in this figure. Cosmic accretion brings mass, thermal energy, turbulent kinetic energy and metal mass into the CGM (light blue arrows on left). The interplay between radiative cooling and the dissipation and pressure support of turbulence in the CGM determines the gas accretion rate into the ISM (light pink arrows on left). The resulting star formation within the galaxy drives feedback in the form of galactic winds (orange arrows) that deposit not only mass and metals back into the CGM but also thermal energy (heating) and kinetic energy (turbulence driving). When the CGM is overpressurized, it can vent mass, energy and metals into the intergalactic medium (pink arrows on right). The inset panel in the bottom right illustrates what we envision for the structure of the CGM in our model: the density and temperature follow assumed radial profiles while the turbulence is characterized by two numbers: the global turbulent velocity of the CGM and the sizes of the largest eddies which together determine the eddy turnover time and hence turbulence dissipation rate.}
\label{fig:structure}
\end{figure*}

Figure \ref{fig:structure} illustrates the essence of our new CGM--galaxy co-evolution model. The model evolves eight state variables associated with the CGM, ISM and stars: the total CGM mass, CGM thermal energy, CGM turbulent kinetic energy, ISM mass, long-lived stellar mass of the central galaxy, and the metal masses of the CGM, ISM and stars. These state variables are evolved according to the following system of coupled ODEs: 

{\small
\begin{eqnarray}\label{eqn:system}
\dot{M}_{\rm CGM} &=& \dot{M}_{\rm in,halo} - \dot{M}_{\rm cool} + \dot{M}_{\rm wind} - \dot{M}_{\rm out,halo}\\
\dot{E}_{\rm CGM}^{\rm th} &=& \dot{E}_{\rm in,halo}^{\rm th} - \dot{E}_{\rm cool} + \dot{E}_{\rm diss} + \dot{E}_{\rm wind}^{\rm th} - \dot{E}_{\rm out,halo}^{\rm th}\\
\dot{E}_{\rm CGM}^{\rm kin} &=& \dot{E}_{\rm in,halo}^{\rm kin} - \dot{E}_{\rm diss} + \dot{E}_{\rm wind}^{\rm kin} - \dot{E}_{\rm out,halo}^{\rm kin}\\
\dot{M}_{\rm ISM} &=& \dot{M}_{\rm cool} - (1-f_{\rm rec})\dot{M}_{\rm SFR} - \dot{M}_{\rm wind}\\
\dot{M}_{\rm star} &=& (1-f_{\rm rec})\dot{M}_{\rm SFR}\\
\dot{M}_{\rm CGM}^{\rm Z} &=& \dot{M}_{\rm in,halo}^{\rm Z} - \dot{M}_{\rm cool}^{\rm Z} + \dot{M}_{\rm wind}^{\rm Z} - \dot{M}_{\rm out,halo}^{\rm Z}\\
\dot{M}_{\rm ISM}^{\rm Z} &=& \dot{M}_{\rm cool}^{\rm Z} +\dot{M}_{\rm yield}^{\rm Z} - (1-f_{\rm rec})\dot{M}_{\rm SFR}^{\rm Z} - \dot{M}_{\rm wind}^{\rm Z}\\
\dot{M}_{\rm star}^{\rm Z} &=& (1-f_{\rm rec})\dot{M}_{\rm SFR}^{\rm Z}
\end{eqnarray}
}
Each of the individual terms in the ODEs has a functional form and associated free parameters that we will now describe in turn. 

\subsection{Cosmic accretion}
The cosmic gas mass accretion rate into halos is 
\begin{equation}\label{eqn:Mdot_in_halo}
\dot{M}_{\rm in,halo} = f_{\rm prev} f_{\rm UV} \fb \dot{M}_{\rm in,gross}
\end{equation}
Here, $\dot{M}_{\rm in,gross}$ is the gross inflow rate of DM and baryonic mass.\footnote{This is different from the \textit{net} inflow rate of mass into the halo which is often estimated by taking the finite difference of the $M_{\rm vir}(t)$ time series. Since our model separately predicts accretion and outflows of gas at the halo radius, we prefer to start with the gross DM inflow rate rather than the net accretion rate.} $f_{\rm b}=0.158$ is the universal baryon fraction from \citet{planck15}. $f_{\rm UV}$ suppresses cosmic accretion below the universal value preferentially in lower mass halos due to photoionization from the cosmic UV background. To estimate $f_{\rm UV}$, we first use Appendix B of \citet{kravtsov04} to compute the redshift-dependent ``filtering halo mass'' at which $50\%$ of baryons are prevented from accreting and then equation (1) of \citet{okamoto08} to compute $f_{\rm UV}$ depending on the ratio of the halo's current mass to the filtering mass. The filtering mass steadily increases from $\sim10^8M_{\odot}$ at $z\sim4$ to $\sim8\times10^9M_{\odot}$ at $z\sim0$ so that $f_{\rm UV}\to0$ for progressively lower mass halos at later times. The $f_{\rm prev}$ parameter accounts for any additional suppression of baryon accretion due to pre-heating, SN feedback, etc. In general, as we will show later, we adopt $f_{\rm prev}\approx1$ for MW-mass halos so that they experience no suppression but $\approx0.3$ for dwarfs such that they accrete only $30\%$ of the universal baryon fraction times the total mass accretion rate.

We assume that halo gas accretion brings in a total energy
\begin{equation}
\dot{E}_{\rm in,halo} = \frac{3}{2}\frac{\kB\Tvir}{\mu\mproton} \dot{M}_{\rm in,halo}
\end{equation}
associated with gas free-falling into the halo. Here $\mu=0.59$ is the mean molecular weight and $T_{\rm vir}$ is the halo virial temperature: 
\begin{equation}
T_{\rm vir} = \frac{1}{2}\frac{\mu m_p}{k_B} V_{\rm vir}^2 \approx 35.9 \left(\frac{V_{\rm vir}}{\rm{km/s}}\right)^2\rm{K}
\end{equation}
with
\begin{equation}
V_{\rm vir} = \sqrt{\frac{G M_{\rm vir}}{R_{\rm vir}}}
\end{equation}
being the virial velocity of the DM halo.

We introduce a free parameter $f_{\rm thermal}^{\rm accretion}$ that partitions this inflowing energy into thermal versus kinetic forms:

\begin{equation}
\dot{E}_{\rm in,halo}^{\rm th} = f_{\rm thermal}^{\rm accretion} \dot{E}_{\rm in,halo}
\end{equation}

\begin{equation}
\dot{E}_{\rm in,halo}^{\rm kin} = (1-f_{\rm thermal}^{\rm accretion}) \dot{E}_{\rm in,halo}
\end{equation}

\subsection{CGM model}
In this subsection we describe the details of how we model the thermal and turbulent structural components of the CGM as well as the gas cooling and accretion rate into the ISM. The relevant CGM structural parameters are illustrated in the inset panel of Figure \ref{fig:structure}.

\subsubsection{Thermal component}
We assume flexible power law profiles for the CGM number density and temperature: 
\begin{eqnarray}
n(r) &=& n_0 \left( \frac{r}{\Rvir} \right)^{\alpha_n} \\
T(r) &=& T_0 \left( \frac{r}{\Rvir} \right)^{\alpha_T}
\end{eqnarray}
where $n_0$ and $T_0$ are the density and temperature respectively at $\Rvir$. The power law slopes $\alpha_n$ and $\alpha_T$ are free parameters -- we typically fix $\alpha_n=-3/2$ and $\alpha_T=0$ corresponding to a steady-state cooling flow solution in an isothermal potential \citep{stern19}, and also very close to what is assumed in our companion paper \citep{carr23}. The normalizations are then derived by requiring that the integrals of the profiles match the state variables $M_{\rm CGM}$ and $E_{\rm CGM}^{\rm th}$ at a given time (the lower limit of $0.1R_{\rm vir}$ is somewhat arbitrary but consistent with our assumed CGM--galaxy boundary for computing fluxes in the FIRE-2 simulations in section \ref{sec:fire}): 

\begin{eqnarray}
\int_{0.1\Rvir}^{\Rvir} 4\pi r^2 \mu m_p n(r) dr &=& M_{\rm CGM}\\
\int_{0.1\Rvir}^{\Rvir} 4\pi r^2 \frac{3}{2}n(r)\kB T(r) dr &=& E_{\rm CGM}^{\rm th}
\end{eqnarray}

These can be solved analytically to get
\begin{equation}\label{eqn:n0}
n_0 = \frac{(3+\alpha_n)M_{\rm CGM}\Rvir^{\alpha_n}}{4\pi\mu\mproton(\Rvir^{3+\alpha_n} - (0.1\Rvir)^{3+\alpha_n})}
\end{equation}
\begin{equation}\label{eqn:T0}
T_0 = \frac{(\alpha_n+\alpha_T+3)\Rvir^{\alpha_n+\alpha_T}E_{\rm CGM}^{\rm th}}{6\pi n_0 \kB (\Rvir^{\alpha_n + \alpha_T + 3} - (0.1\Rvir)^{\alpha_n+\alpha_T+3})}
\end{equation}
Note that, contrary to the assumptions of traditional SAMs, $T_0$ is generally not equal to the halo virial temperature $T_{\rm vir}$. In principle, we could also similarly set up profiles for the CGM metallicity and turbulent velocity. However for simplicity we assume that the CGM metallicity is constant with radius. We also adopt a single global characteristic turbulent velocity since introducing a radial dependence for $v_{\rm turb}$ would require additional complications in order to relate it to our assumed model for the circular velocity profile of the underlying DM halo.

Then the radiative cooling rate for the thermal component is 
\begin{equation}\label{eqn:Edotcool}
\dot{E}_{\rm cool} = \int_{0.1\Rvir}^{\Rvir}4\pi r^2 n(r)^2 \Lambda(r) dr
\end{equation}
where $\Lambda(r)$ is the cooling function. We adopt the \citet{wiersma09} cooling tables which assume collisional ionization equilibrium (CIE) plus photoionization from the \citet{haardt01} UV background, thus causing $\Lambda$ to depend on density, temperature, metallicity and redshift. The cooling tables only go up to $z=8.989$ -- above this redshift (or for any other extrapolation) we set $\Lambda=0$. These cooling tables include photoionization heating of gas that is already in the CGM whereas our prescription for $f_{\rm UV}$ above inhibits the accretion of gas into the halo in the first place \citep[see also][]{benson02}. Note that our model does not account for the possibility that cold clouds condense out of the warm/hot phase thereby lowering its density and cooling rate, nor for the scenario that these cold clouds can contribute energy and mass back to the warm/hot phase as they free-fall towards the galaxy \citep{maller04,murray04,forbes19b,fielding20}.

\subsubsection{Turbulent kinetic component}
The turbulent kinetic energy of the CGM is assumed to dissipate on a timescale \citep{stone98,maclow99b}
\begin{equation}\label{eqn:tturb}
t_{\rm turb} = \frac{R_{\rm turb}}{v_{\rm turb}}
\end{equation}
so that 
\begin{equation}
\dot{E}_{\rm diss} = \frac{E_{\rm kin}}{t_{\rm turb}}
\end{equation}
$t_{\rm turb}$ is known as the characteristic eddy turnover time, $R_{\rm turb}$ is the characteristic size of the largest turbulent eddies, and $v_{\rm turb}$ is their corresponding turbulent velocity (this can also be thought of as the specific turbulent kinetic energy of the CGM). We take
\begin{equation}\label{eqn:vturb}
v_{\rm turb} = \sqrt{\frac{2 E_{\rm CGM}^{\rm kin}}{M_{\rm CGM}}}
\end{equation}
We do not have a priori knowledge of what $R_{\rm turb}$ should be and how it may vary with halo mass, redshift and/or CGM properties. Thus we make the simple assumption that 
\begin{equation}
R_{\rm turb}(t) \propto R_{\rm vir}(t)
\end{equation}
In other words, we assume that the sizes of the largest turbulent eddies in the CGM are proportional to the halo virial radius and that this size sets the timescale on which the turbulent cascade proceeds. The proportionality factor may or may not be a function of time depending on the driving sources of the turbulence. In Section \ref{sec:Rturb} we will present our fiducial function that relates $R_{\rm turb}$ and $R_{\rm vir}$ as a function of time. 


\subsubsection{CGM mass cooling and inflow rate}
So far we have dealt with CGM energetics, but now we must consider the rate at which the CGM loses mass via accretion into the ISM. We compute the ISM accretion rate as 
\begin{equation}\label{eqn:mdot_cool}
\dot{M}_{\rm cool} = \frac{M_{\rm CGM}}{t_{\rm cool,eff} + t_{\rm ff,eff}}
\end{equation}
Here, $t_{\rm cool,eff}$ is the ``effective'' timescale on which the CGM would radiate away its thermal energy given the net cooling rate (equation \ref{eqn:Edotcool}) minus the turbulence dissipation rate: 
\begin{equation}\label{eqn:tcooleff}
t_{\rm cool,eff} = \frac{E_{\rm CGM}^{\rm th}}{\dot{E}_{\rm cool}-\dot{E}_{\rm diss}}
\end{equation}
There is another limiting timescale for ISM accretion that we call the ``effective'' free-fall time: 
\begin{equation}\label{eqn:tffeff}
t_{\rm ff,eff} = \frac{R_{\rm max}}{V_{\rm max}} \left( 1 + \frac{v_{\rm turb}^2}{V_{\rm max}^2} \right)^{1/2}
\end{equation}
After CGM gas has had sufficient time to cool (or while it is cooling), it will take this additional amount of time to free-fall into the galaxy from some characteristic CGM radius which we take to be $R_{\rm max}\approx2.16R_{\rm s}$.\footnote{It is common practice to use the halo dynamical time $R_{\rm vir}/V_{\rm vir}$ which can be much longer than $R_{\rm max}/V_{\rm max}$. However we find that this is generally too long in our new model because the CGM ends up losing thermal energy to radiative cooling much faster than it loses mass via accretion into the ISM. This causes the CGM specific energy to become quite low and prevents halo outflows via our new CGM overpressurization mechanism described in subsection \ref{sec:overpressurization} below.} This is the radius where the halo circular velocity profile reaches its maximum with $R_{\rm s}=R_{\rm vir}/c$ being the scale radius of an NFW halo with concentration $c$. Thus in the absence of turbulent pressure support, the cooling gas would inflow on a timescale $t_{\rm ff}=R_{\rm max}/V_{\rm max}$. However, Equation \ref{eqn:tffeff} multiplies this by a smooth function that increases the inflow timescale when the ratio of the turbulent pressure support ($\sim\rho v_{\rm turb}^2$) to the local gravitational potential energy density ($\sim \rho V_{\rm max}^2$) is large, and reverts to the baseline free-fall time otherwise. The exact functional form is arbitrary but achieves our desired effect in a smooth manner. To estimate $V_{\rm max}$, we use an analytic approximation for the circular velocity profile of an NFW halo with a given concentration and $V_{\rm vir}$ \citep[equation 11.26 from][with $x\equiv r/R_{\rm vir}$]{mvdbw10}: 
\begin{equation}
V_{\rm circ}(r) = V_{\rm vir} \left(\frac{1}{x}\frac{\ln(1+cx)-cx/(1+cx)}{\ln(1+c) - c/(1+c)} \right)^{1/2}
\end{equation}

Another way to compute $\dot{M}_{\rm cool}$ would be to radially integrate over $4\pi r^2 \rho(r) / (t_{\rm cool,eff}(r)+t_{\rm ff,eff}(r))$. Alternatively, we could compute a ``cooling radius'' $R_{\rm cool}$ within which $t_{\rm cool,eff}< t_{\rm ff,eff}$ and then set $\dot{M}_{\rm cool}=M_{\rm CGM}(R_{\rm cool}/R_{\rm vir})/t_{\rm ff,eff}$ following the approach originally laid out by \citet[][]{whitefrenk91} and later adopted by most SAMs. While both of these methods would naturally take into account the fact that cooling and free-fall times are longer at larger radii and negate the need for arbitrarily choosing a single characteristic free-fall radius, it would introduce additional complications for parameterizing a $v_{\rm turb}(r)$ profile and relating it to the underlying $V_{\rm circ}(r)$ profile of the DM halo to compute $t_{\rm ff,eff}(r)$. Instead here we adopt the simpler approach of treating the CGM cooling and free-fall times with single global parameters. 

\subsection{Star formation and supernova-driven winds}
We model the small-scale physics of star formation and SN-driven winds in a very simple, flexible way. The star formation rate is modeled using a single depletion time parameter for the entire ISM mass: 
\begin{equation}
\dot{M}_{\rm SFR} = \frac{M_{\rm ISM}}{t_{\rm dep}}
\end{equation}
The SN-driven mass outflow rate is simply
\begin{equation}
\dot{M}_{\rm wind} = \eta_{\rm M} \dot{M}_{\rm SFR}
\end{equation}
where $\eta_{\rm M}$ is the wind mass loading factor. The corresponding energy outflow rate of the SN-driven wind is 
\begin{equation}
\dot{E}_{\rm wind} = \dot{M}_{\rm wind} v_{\rm B}^2
\end{equation}
where 
\begin{equation}\label{eqn:vB}
v_{\rm B}^2 = \frac{1}{2}v^2 + \frac{3}{2}c_s^2
\end{equation}
is the Bernoulli velocity that quantifies the specific energy of the wind. Here $v$ is the wind velocity and $c_s$ is the wind sound speed corresponding to its temperature. Note that 
\begin{equation}
\eta_{\rm E} = \frac{\dot{E}_{\rm out}}{e_{\rm SN}\rm{SFR}} = \frac{\dot{M}_{\rm out}v_{\rm B}^2}{e_{\rm SN}\rm{SFR}} = \eta_{\rm M}\frac{v_{\rm B}^2}{e_{\rm SN}}
\end{equation}
where $e_{\rm SN}=10^{51}$erg/(100$M_{\odot}$) is the SN energy produced per $100M_{\odot}$ of stars formed, consistent with the \citet{kroupa01} initial mass function (IMF). Thus we will use $v_{\rm B}$ and $\eta_{\rm E}/\eta_{\rm M}$ interchangeably to refer to the wind specific energy. Finally we introduce a free parameter $f_{\rm thermal}^{\rm wind}$ that partitions the outflowing wind energy into thermal and kinetic forms (analogous to how we partitioned the cosmic accretion energy):
\begin{eqnarray}
\dot{E}_{\rm wind}^{\rm th} &=& f_{\rm thermal}^{\rm wind}\dot{E}_{\rm wind}\\
\dot{E}_{\rm wind}^{\rm kin} &=& (1-f_{\rm thermal}^{\rm wind})\dot{E}_{\rm wind}
\end{eqnarray}
It follows from Equation \ref{eqn:vB} that $f_{\rm thermal}^{\rm wind}=3/4$ corresponds to a Mach number of one. Various functional forms can be adopted for the free parameters $t_{\rm dep}$, $\eta_{\rm M}$, $v_{\rm B}$ (or equivalently $\eta_{\rm E}$) and $f_{\rm thermal}^{\rm wind}$. We will present our fiducial functional forms in Section \ref{sec:fire}.

\subsection{CGM over-pressurization and halo outflows}\label{sec:overpressurization}
The energy pumped into the CGM from SN winds and cosmic accretion may over-pressurize it compared to its own binding energy\footnote{This is an overly simple approximation for the CGM binding energy since gas in the inner halo will be more tightly bound to the halo than gas at large radii.}
\begin{equation}
E_{\rm bind} = \frac{3}{2}\frac{\kB\Tvir}{\mu\mproton} M_{\rm CGM}
\end{equation}
This over-pressurization may happen if radiative cooling cannot keep up with heating and/or if the turbulent energy is not dissipating fast enough. In this case we envision that some fraction of the CGM will become unbound from the halo and drive outflows at $\Rvir$ to naturally decrease the level of over-pressurization \citep[see also][]{carr23}

The total excess energy outflow rate at $\Rvir$ is
\begin{equation}
\dot{E}_{\rm out,halo} = \frac{\max(E_{\rm CGM}^{\rm th}+E_{\rm CGM}^{\rm kin}-E_{\rm bind}, 0)}{t_{\rm dyn,halo}}
\end{equation}
where $t_{\rm dyn,halo}=\Rvir/\Vvir$ as usual.\footnote{Alternatively we could have the turbulent component flow out of the halo on a timescale $R_{\rm vir}/v_{\rm turb}$ and the thermal energy flow out on a timescale $R_{\rm vir}/c_{\rm s}$, but we avoid this complication.} For the thermal versus kinetic energy partitioning of the halo outflows, $f_{\rm thermal}^{\rm out}$, we assume the current thermal energy support fraction of the CGM (note that this is not a free parameter):
\begin{equation}
f_{\rm thermal}^{\rm CGM} = E_{\rm CGM}^{\rm th} / (E_{\rm CGM}^{\rm th}+E_{\rm CGM}^{\rm kin})
\end{equation}
so that 
\begin{eqnarray}
\dot{E}_{\rm out,halo}^{\rm th} &=& f_{\rm thermal}^{\rm CGM} \dot{E}_{\rm out,halo}\\
\dot{E}_{\rm out,halo}^{\rm kin} &=& (1-f_{\rm thermal}^{\rm CGM}) \dot{E}_{\rm out,halo}
\end{eqnarray}

Finally the mass outflow rate at $R_{\rm vir}$ is just the total energy outflow rate divided by the specific energy of the halo outflows, $e_{\rm out,halo}$, which by default we take equal to the specific energy of the CGM:
\begin{equation}
\dot{M}_{\rm out,halo} = \frac{\dot{E}_{\rm out,halo}}{E_{\rm CGM}/M_{\rm CGM}}
\end{equation}
Because we are forcing the halo energy outflows to have the same thermal vs. kinetic split as the CGM, halo outflows do not change the CGM temperature. However this can be modified by introducing another multiplicative free parameter to rescale the denominator to much higher or lower specific energies relative to the CGM.

\subsection{Chemical evolution}
We use the instantaneous recycling approximation to track the production and flow of all metals combined (i.e., we do not track different species or production channels). We follow section 10.4.2 of \citet{mvdbw10} which is exactly equivalent to the ODEs implemented in some SAMs \citep{cole00,delucia04,lagos18}. Following \citet{tinsley80}, we assume that new stars form with the same metallicity as the ISM ($Z_{\rm ISM}$) so that 
\begin{equation}
\dot{M}_{\rm SFR}^{\rm Z} = Z_{\rm ISM} \dot{M}_{\rm SFR}
\end{equation}
Note that equations (7) and (8) already include the $(1-f_{\rm rec})$ factor to account for the instantaneous return fraction of ISM metals via stellar winds and SNe. The addition of new metals to the ISM from stellar nucleosynthesis is given by
\begin{equation}\label{eqn:MZyield}
\dot{M}_{\rm yield}^{\rm Z} = (1-f_{\rm rec})y_Z \dot{M}_{\rm SFR}
\end{equation}
The parameter $y_Z$ is defined as the ratio of new metal mass ejected from stars divided by the total mass locked up in low-mass stars and stellar remnants. Thus, by definition, $y_Z$ includes a factor of $(1-f_{\rm rec})$ in its denominator \citep[see equation 10.120 of][]{mvdbw10}. Note that some SAMs define $y_Z$ differently by excluding the factor of ($1-f_{\rm rec}$) in its denominator and then also removing it from equation \ref{eqn:MZyield} \citep{somerville15,lagos18} but this is identical to our approach. We assume $(1-f_{\rm rec})y_Z\approx0.02$ consistent with a \citet{kroupa01} IMF.

The metal accretion rate into the halo is assumed to be 
\begin{equation}
\dot{M}_{\rm in,halo}^{\rm Z} = Z_{\rm in,halo} \dot{M}_{\rm in,halo}
\end{equation}
where $Z_{\rm in,halo}$ is a free parameter for the inflow metallicity. For pristine accretion, this would be $Z_{\rm in,halo}=0$ although we expect some pre-enrichment from recycling of galactic winds and/or outflows from nearby halos. Note that some SAMs model a separate ``ejected'' component into which they deposit winds and metals ejected from halos, and then compute the recycling rate of this ejected gas back into the halo \citep{henriques13,white15}. However this introduces additional complications such as wind escape fractions, mass-dependent recycling times, and contribution of additional metals beyond those produced by the central galaxy that we do not include in this work.

The metal accretion rate into the ISM from CGM cooling is taken to be proportional to the CGM metallicity:
\begin{equation}
\dot{M}_{\rm cool}^{\rm Z} = Z_{\rm CGM} \dot{M}_{\rm cool}
\end{equation}
This assumes perfect mixing of inflows and outflows in the CGM and ignores the possibility that metal-enriched material is expected to cool more efficiently \citep{hobbs15}.

The metal mass carried by SN-driven winds is assumed to be proportional to the ISM metallicity:
\begin{equation}
\dot{M}_{\rm wind}^{\rm Z} = Z_{\rm ISM}\dot{M}_{\rm wind}
\end{equation}
In principle the wind metallicity could be different from the ISM metallicity if, for example, there was minimal entrainment and the wind contained mostly SN ejecta. This would require introducing a new wind enrichment factor or metal loading factor \citep[e.g.,][]{kim20,sharda21,carr23} but we defer this complication to future work. Finally, we assume that the metal outflow rate from the halo is directly proportional to the CGM metallicity:
\begin{equation}
\dot{M}_{\rm out,halo}^{\rm Z} = Z_{\rm CGM} \dot{M}_{\rm out,halo}
\end{equation}

\subsection{Summary of free parameters}
Table \ref{tab:params} lists all of our model parameters and whether they are fixed or allowed to vary. In total our model has 16 parameters governing cosmic accretion, CGM structure, star formation and stellar feedback, and chemical evolution. However, most of these parameters except one ($R_{\rm turb}$) are either fixed based on simple physical arguments or on our calibration to the FIRE-2 simulations (see section \ref{sec:fire}). In many cases, a parameter is not simply a constant but rather is given by a functional form with a number of additional arguments that quantify the redshift and/or halo mass dependence of that parameter. In these cases, we refer the reader to the relevant sections quoted in the table for the exact parameterizations. 

\begin{table*}\scriptsize
\centering
\begin{tabular}{ l l l l }
 Parameter & Fixed/Free & Value & Meaning \\\hline\hline
 Cosmic accretion & & & Sections 2.1, 4.3.3, 4.3.4 \\
 $f_{\rm UV}$ & Fixed & \citet{okamoto08} & Suppression factor for cosmic accretion due to UV background \\
 $f_{\rm prev}$ & Fixed* & Logistic & Suppression factor for cosmic accretion due to preventative feedback \\
 $f_{\rm thermal}^{\rm accretion}$ & Fixed* & Power law & Fraction of accretion energy that is thermal rather than kinetic \\\hline
 CGM structure & & & Sections 2.2 and 4.3.5 \\
 $\alpha_n$ & Fixed & -1.5 & Slope of CGM density power law \\  
 $\alpha_T$ & Fixed & 0.0 & Slope of CGM temperature power law \\
 $R_{\rm turb}$ & Free & Logistic & Sizes of largest CGM eddies (controls turbulence dissipation rate) \\
 $r_{\rm ff}$ & Fixed & $R_{\rm max}$ & Radius at which to compute the effective free-fall time of the CGM \\\hline
 Star formation and stellar feedback & & & Sections 2.3, 2.4 and 4.3.1 \\
 $t_{\rm dep}$ & Fixed* & Power law & ISM depletion time \\
 $\eta_{\rm M}$ & Fixed* & Power law & Mass loading of SN-driven winds \\
 $v_{\rm B}$ & Fixed* & Power law & Specific energy of SN-driven winds \\
 $f_{\rm thermal}^{\rm wind}$ & Fixed* & Logistic & Fraction of SN wind energy that is thermal rather than kinetic \\
 $f_{\rm thermal}^{\rm out}$ & Fixed & $f_{\rm thermal}^{\rm CGM}$ & Fraction of halo outflow energy that is thermal rather than kinetic \\
 $e_{\rm out,halo}$ & Fixed & $E_{\rm CGM}/M_{\rm CGM}$ & Specific energy of halo outflows \\\hline
 Chemical evolution & & & Sections 2.5 and 4.3.2 \\
 $f_{\rm rec}$ & Fixed & 0.4 & Instantaneous stellar recycling fraction \\
 $(1-f_{\rm rec})y_{\rm Z}$ & Fixed & 0.02 & Nucleosynthetic yield from star formation and stellar evolution \\
 $Z_{\rm in,halo}$ & Fixed* & Power law & Metallicity of cosmic accretion \\
\end{tabular}
\caption{\label{tab:params}List of model parameters. Fixed* means the parameter was fixed based on FIRE-2. If the value is not a constant and is instead given as a reference or functional form, then there are a number of additional arguments that control the redshift and/or halo mass dependence of the parameter (see the relevant section for the exact parameterization). In total we have 16 parameters but effectively all of these except one ($R_{\rm turb}$) are fixed based on the FIRE-2 simulations or simple physical arguments.}
\end{table*}

\subsection{Numerical details}\label{sec:numerics}
We solve the system of ODEs defined in Equation \ref{eqn:system} using the Python \texttt{scipy.integrate.solve\_ivp} ODE solver. Specifically, we use the adaptive-timestep implicit method ``BDF'' (backward differentiation formula) which gives identical results compared to the adaptive-timestep explicit methods ``RK23'' and ``RK45'' (Runge-Kutta) but with fewer overall iterations required. Our results are insensitive to the choice of initial conditions for the state variables as long as they are reasonably small and close to, but not exactly, zero (this is because the very early evolution is driven by the rapid cosmic assembly of halos).

Many of the individual terms in our ODEs require external time-dependent inputs related to the halo assembly history. These include the halo mass, halo radius, halo concentration and gross DM accretion rate. Since we use measurements of these halo properties from numerical simulations (see Section \ref{sec:fire}) and since those measurements can be noisy (the typical spacing between simulation outputs is $\sim2-10$ Myr), we need to smooth and interpolate those input time series with a sufficiently high order function to ensure optimally adaptive timestepping and stability for our ODE solver. For this, we make use of \texttt{scipy.interpolate.UnivariateSpline} to fit a smoothing spline of degree 5 with a smoothing factor of 2. Furthermore, when comparing our predicted time series for any given quantity to the measurements of that property from simulations, we smooth the simulation time series with a Gaussian whose standard deviation is 10 (in units of number of adjacent data points for simplicity) using \texttt{scipy.ndimage.gaussian\_filter1d}. This is done because our model is based on ODEs which are inherently smooth and do not account for time delays and stochasticity.

\section{Equilibrium Behavior of the Model}\label{sec:eq}
In this section, we illustrate the equilibrium behavior of our ODEs by varying key state variables and model parameters at a single redshift. This instantaneous exercise will help us better understand the fully time-dependent predictions of our model when we run it on halo assembly histories in the next section. 

\subsection{Identifying model equilibria in $v_{\rm turb}-T_0$ space}
$T_0$ and $v_{\rm turb}$ respectively quantify the specific thermal energy and specific kinetic energy of the CGM in our model (see Equations \ref{eqn:T0} and \ref{eqn:vturb}, respectively). These two parameters hence reflect the amount of cooling, heating and overpressurization in the CGM, which in turn will affect star formation and supernova feedback. Since these two parameters are also novel compared to previous SAMs (which fix $T_0=T_{\rm vir}$ and neglect CGM turbulence), it is important to understand their equilibrium values (where $\dot{E}_{\rm th}=0$ and $\dot{E}_{\rm kin}=0$) since our model will tend to evolve towards those values.

As an illustrative example, we adopt model parameters that are reasonable for a Milky Way mass halo at $z=0$: 
\begin{enumerate}
\item $M_{\rm vir}=10^{12}M_{\odot}$, $R_{\rm vir}=275$ kpc, $c_{\rm NFW}=15$. These together give $V_{\rm vir}=125$ km/s and $T_{\rm vir}=5.6\times10^5$ K.
\item $\dot{M}_{\rm in,gross}=75M_{\odot}$/yr
\item $M_{\rm CGM}=5\times10^{10}M_{\odot}$ and $Z_{\rm CGM}=0.3Z_{\odot}$
\item $\alpha_n=-3/2$ and $\alpha_T=0.0$
\item $R_{\rm turb}=0.5R_{\rm vir}$
\item $f_{\rm thermal}^{\rm accretion}=0.5$ and $f_{\rm thermal}^{\rm wind}=0.5$
\item $\eta_{\rm M}=0.5$ and $v_{\rm B}=300$ km/s which together imply $\eta_{\rm E}=\eta_{\rm M}v_{\rm B}^2/e_{\rm SN}\approx0.1$ 
\item $e_{\rm out,halo} = e_{\rm CGM}$
\end{enumerate}

Then we set up a 2D grid spanning a range of $v_{\rm turb}$ and $T_0$ values. At each point in this grid, we re-compute $E_{\rm kin}=\frac{1}{2}M_{\rm CGM}v_{\rm turb}^2$ and $E_{\rm th} = \frac{3}{2} M_{\rm CGM} k_B T_0 / (\mu m_p)$, which in turn will change the instantaneous $\dot{E}_{\rm diss}$, $\dot{E}_{\rm cool}$ and $\dot{M}_{\rm cool}$. Since we also want to see the effect of changing $v_{\rm turb}$ and $T_0$ on the SFR and winds, we re-compute $\rm{SFR}=\dot{M}_{\rm cool}$. This is an approximation because we are ignoring the ODEs linking $\dot{M}_{\rm cool}$ to $\dot{M}_{\rm ISM}$ and hence not self-consistently predicting the SFR. However, setting the SFR equal to the ISM accretion rate should be appropriate for our case of a MW-mass halo at $z=0$ (more generally, in lower mass halos and high-redshift MW progenitors, we expect the SFR to only be a small fraction of the ISM accretion rate). Finally, we are then in a position to compute $\dot{E}_{\rm th}$ and $\dot{E}_{\rm kin}$ across our grid.

\begin{figure*}
\centering
\includegraphics[width=0.9\hsize]{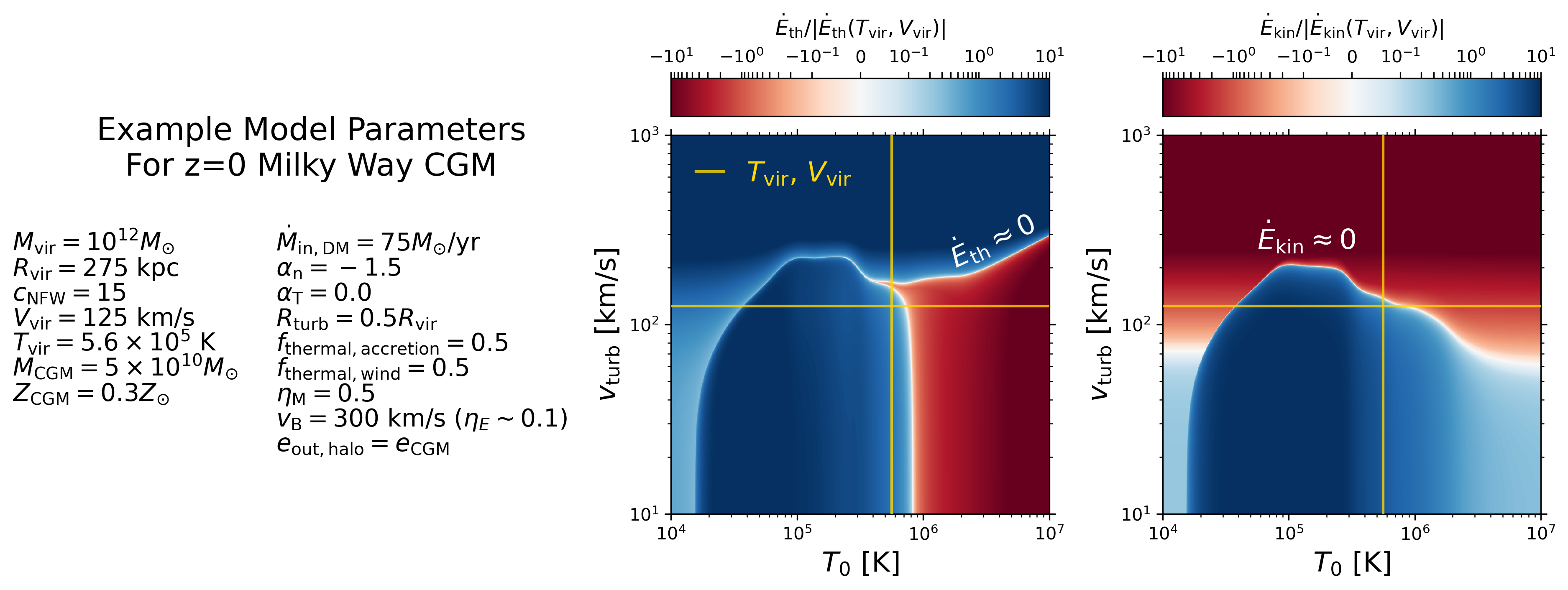}
\caption{Identifying equilibrium values of $T_0$ (middle) and $v_{\rm turb}$ (right) using parameters characteristic for a MW-mass halo at $z=0$ (values given on left). The colorbars show $\dot{E}_{\rm th}$ and $\dot{E}_{\rm kin}$ normalized by their values for $T_0=T_{\rm vir}$ and $v_{\rm turb}=V_{\rm vir}$. Both $T_0$ and $v_{\rm turb}$ have narrow strips of possible equilibrium values and these equilibria are generally stable: perturbations away from the white strips will push the model back towards equilibrium. The exception is low temperature solutions ($T_0\lesssim5\times10^5$ K) around which  $\dot{E}_{\rm th}>0$. The point where the equilibrium regions for the two panels intersect gives the common equilibrium solution that the full model will tend to evolve toward. For our choice of $z=0$ MW parameters, this intersection happens when $T_0\approx8\times10^5$ K (slightly hotter than $T_{\rm vir}$) and $v_{\rm turb}\approx130$ km/s (comparable to $V_{\rm vir}$).}
\label{fig:heatmap}
\end{figure*}

Figure \ref{fig:heatmap} illustrates how $\dot{E}_{\rm th}$ and $\dot{E}_{\rm kin}$ vary across our $v_{\rm turb}-T_0$ grid. The narrow strips where $\dot{E}_{\rm th}=0$ and $\dot{E}_{\rm kin}=0$ respectively pinpoint the range of possible equilibrium values of $T_0$ and $v_{\rm turb}$ for our choice of $z=0$ MW-like parameters. The equilibria are generally stable: for small perturbations away from the strips, the model will be pushed back toward equilibrium. The exception is low temperature solutions ($T_0\lesssim5\times10^5$ K) around which $\dot{E}_{\rm th}>0$ indicating that those equilibria are unstable and that small perturbations would drive the model to a hotter temperature. The intersection of the two equilibrium regions for both $\dot{E}_{\rm th}=0$ and $\dot{E}_{\rm kin}=0$ gives the combination of equilibrium $T_0$ and $v_{\rm turb}$ that the full model will tend to evolve toward (and once reaching these values, the model will stay there except for forcing terms from cosmological accretion). For our choice of $z=0$ model parameters, this intersection happens at $T_0\approx8\times10^5$ K (slightly hotter than $T_{\rm vir}$) and $v_{\rm turb}\approx130$ km/s (comparable to $V_{\rm vir}$). 

\subsection{How do equilibria depend on model parameters?}\label{sec:eq1D}
Having illustrated the existence of equilibrium solutions in our model, a natural follow-up question is to ask how these equilibria depend on our choice of model parameters. Figure \ref{fig:eq1D} is similar to Figure \ref{fig:heatmap} but now we vary a few model parameters alongside $T_0$ while fixing $v_{\rm turb}=V_{\rm vir}$ to see the effect on $\dot{E}_{\rm th}$, and then repeat the variation alongside $v_{\rm turb}$ while fixing $T_0=T_{\rm vir}$ to see the effect on $\dot{E}_{\rm kin}$. The two model parameters that we explore are the wind specific energy and the largest eddy turnover scale. We expect our model to be quite sensitive to these two parameters (in addition to redshift, halo mass, CGM metallicity, etc. but we leave those for section \ref{sec:eqevol} where we will show how model equilibria evolve naturally along individual halo assembly histories). The following subsections examine each of these parameter variations in turn.

\subsubsection{Wind specific energy} 
We vary the wind specific energy between $\eta_{\rm E}/\eta_{\rm M}=0.01-1.0$ while keeping the wind mass loading factor fixed at our fiducial value of $\eta_{\rm M}=0.5$. Note that $\eta_{\rm E}/\eta_{\rm M}=0.01$ implies $v_{\rm B}=\sqrt{0.01 e_{\rm SN}}\approx70$ km/s and $\eta_{\rm E}/\eta_{\rm M}=1.0$ implies $v_{\rm B}=\sqrt{e_{\rm SN}}\approx700$ km/s.

The top-left panel of Figure \ref{fig:eq1D} shows that as the wind specific energy increases, the equilibrium value of $T_0$ rises. This makes sense because it becomes harder for CGM cooling to keep up with the extra heating. On the other hand, decreasing the wind specific energy leads to a drop in $T_0$ because the winds deposit cold mass in the CGM without a commensurate increase in its thermal energy, thus enabling more cooling. Interestingly, there is a narrow region around $\eta_{\rm E}/\eta_{\rm M}\approx0.1$ where the temperature changes quite abruptly. This likely reflects atomic cooling physics: at lower temperatures, the CGM would be on the thermally unstable part of the cooling curve so extra heating may only gradually increase the average CGM temperature. As the model approaches the peak of the cooling curve at $T_0\sim10^5$ K, any extra heating cannot be compensated by an increase in cooling and this quickly drives the model to a hotter equilibrium temperature. It is not a coincidence that the model equilibrium itself is unstable near this thermal instability region where $T_0\approx3\times10^5$ and $\eta_{\rm E}/\eta_{\rm M}\approx0.1$. 

The top-right panel of Figure \ref{fig:eq1D} shows qualitatively similar behavior for the dependence of equilibrium $v_{\rm turb}$ on $\eta_{\rm E}/\eta_{\rm M}$. As SN winds carry less kinetic energy into the CGM (with mass loading fixed), the equilibrium $v_{\rm turb}$ will naturally be lower since there is less turbulence driving and vice versa. It is interesting that the equilibrium $v_{\rm turb}$ varies more slowly and over a smaller range than $T_0$. This is likely caused by the fact that the turbulence dissipation rate goes as $\dot{E}_{\rm diss}=E_{\rm kin}/t_{\rm turb}\propto v_{\rm turb}^3$ and we are not independently varying the dissipation timescale, so when there is a lot of turbulence it will decay quickly and the equilibrium $v_{\rm turb}$ will not rise indefinitely. We will show next that the equilibrium $v_{\rm turb}$ is more sensitive to the largest eddy turnover scale that we assume in our model.

\subsubsection{Largest eddy turnover scale}
\begin{figure}
\centering
\includegraphics[width=0.5\hsize]{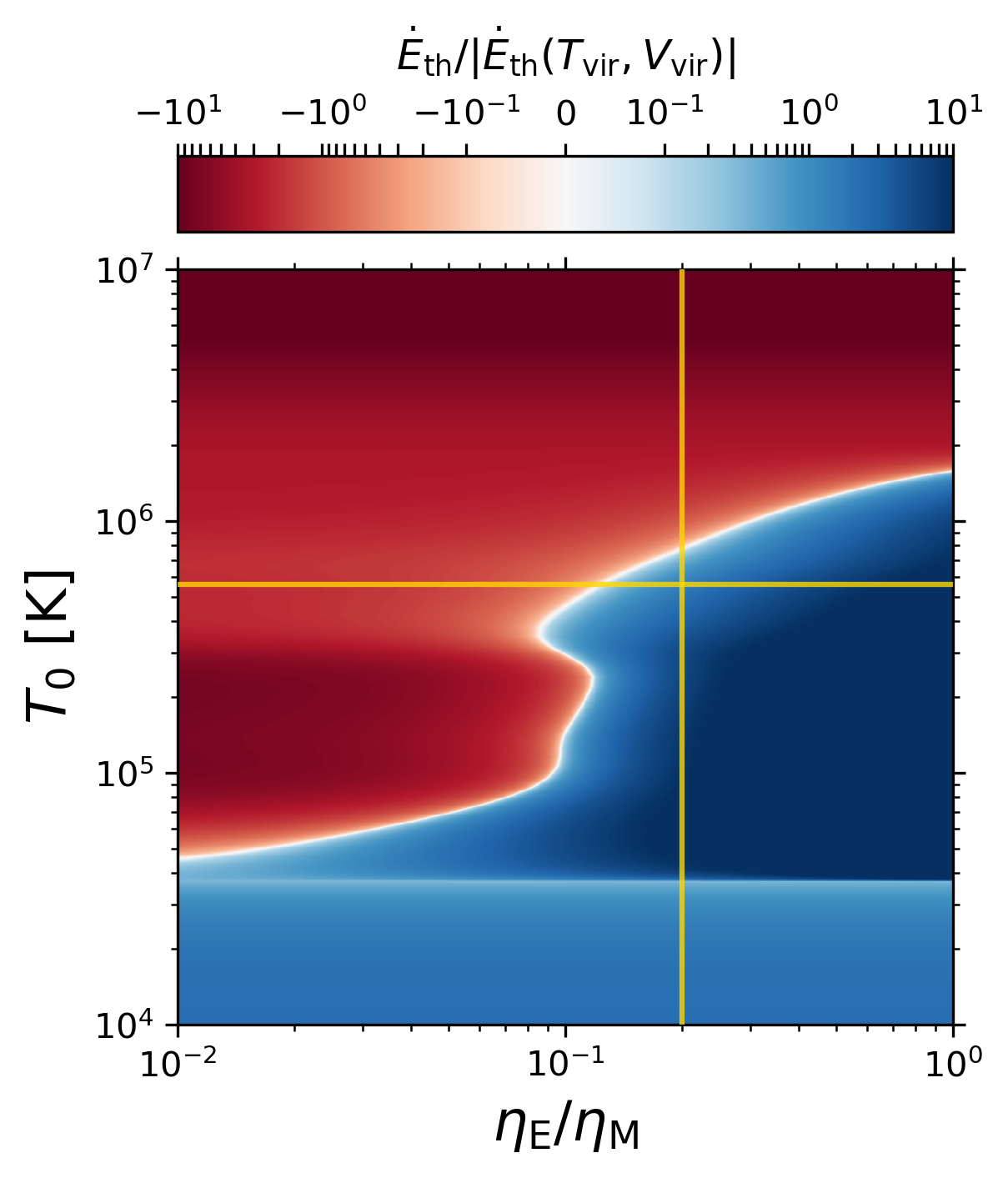}\includegraphics[width=0.5\hsize]{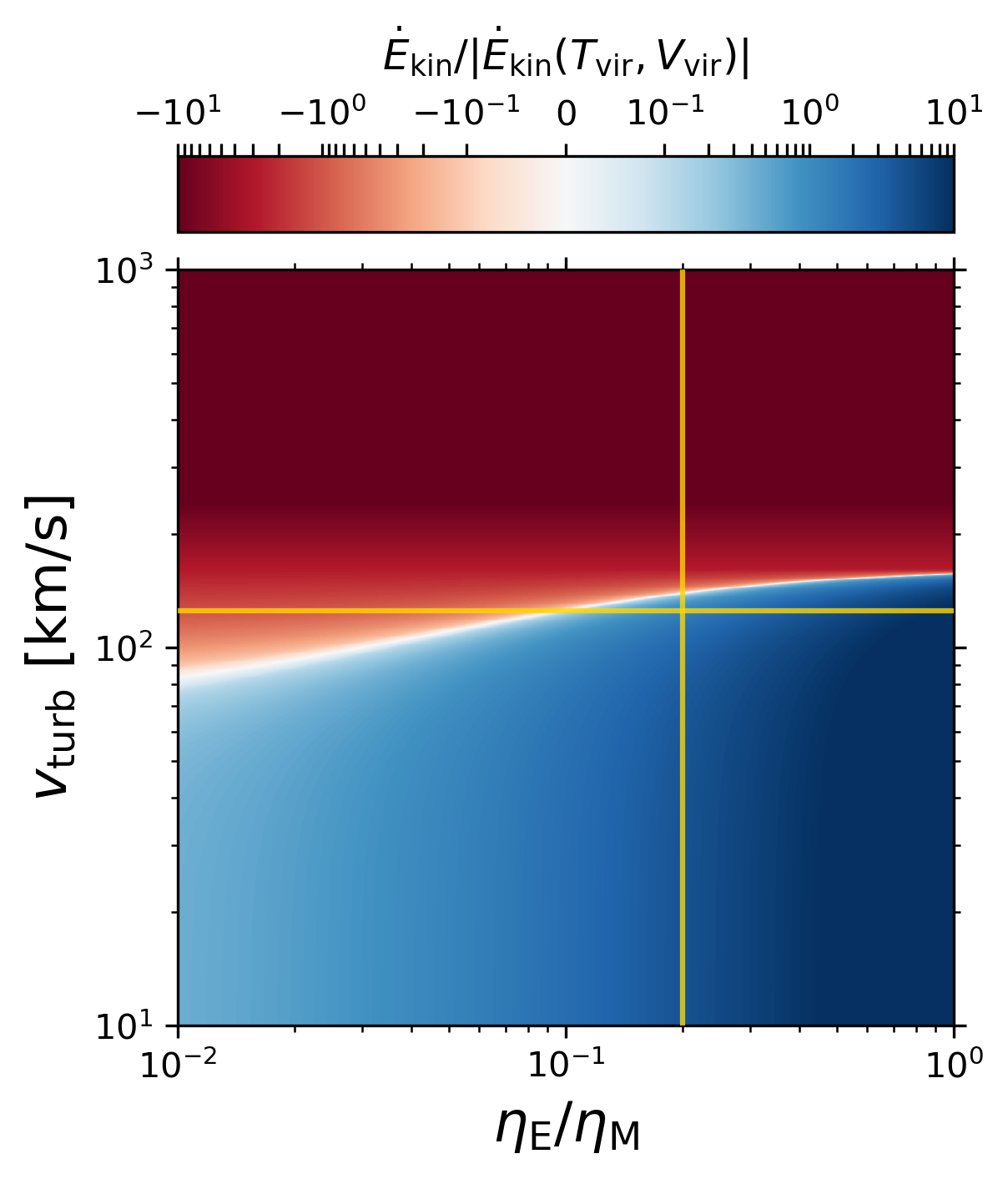}

\includegraphics[width=0.5\hsize]{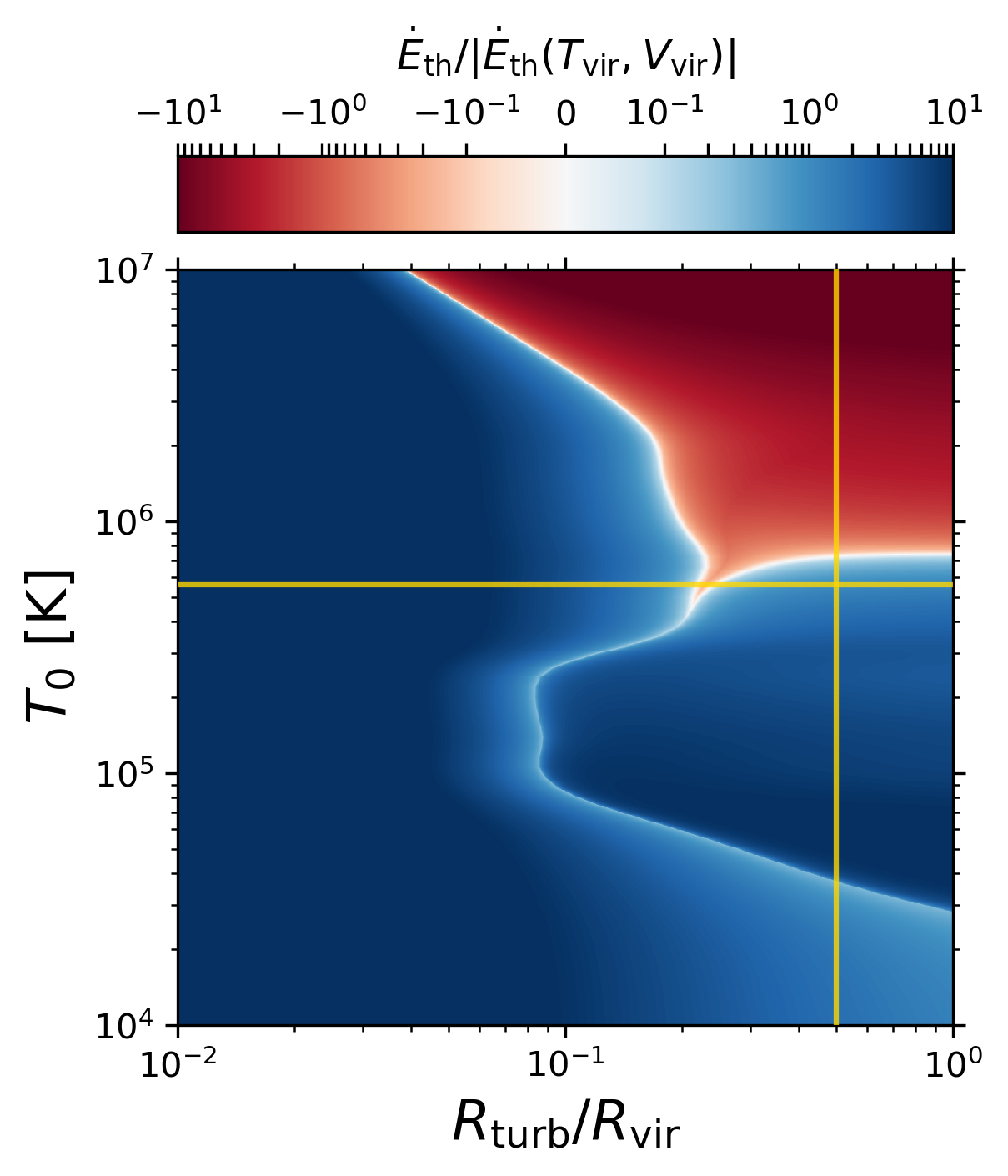}\includegraphics[width=0.5\hsize]{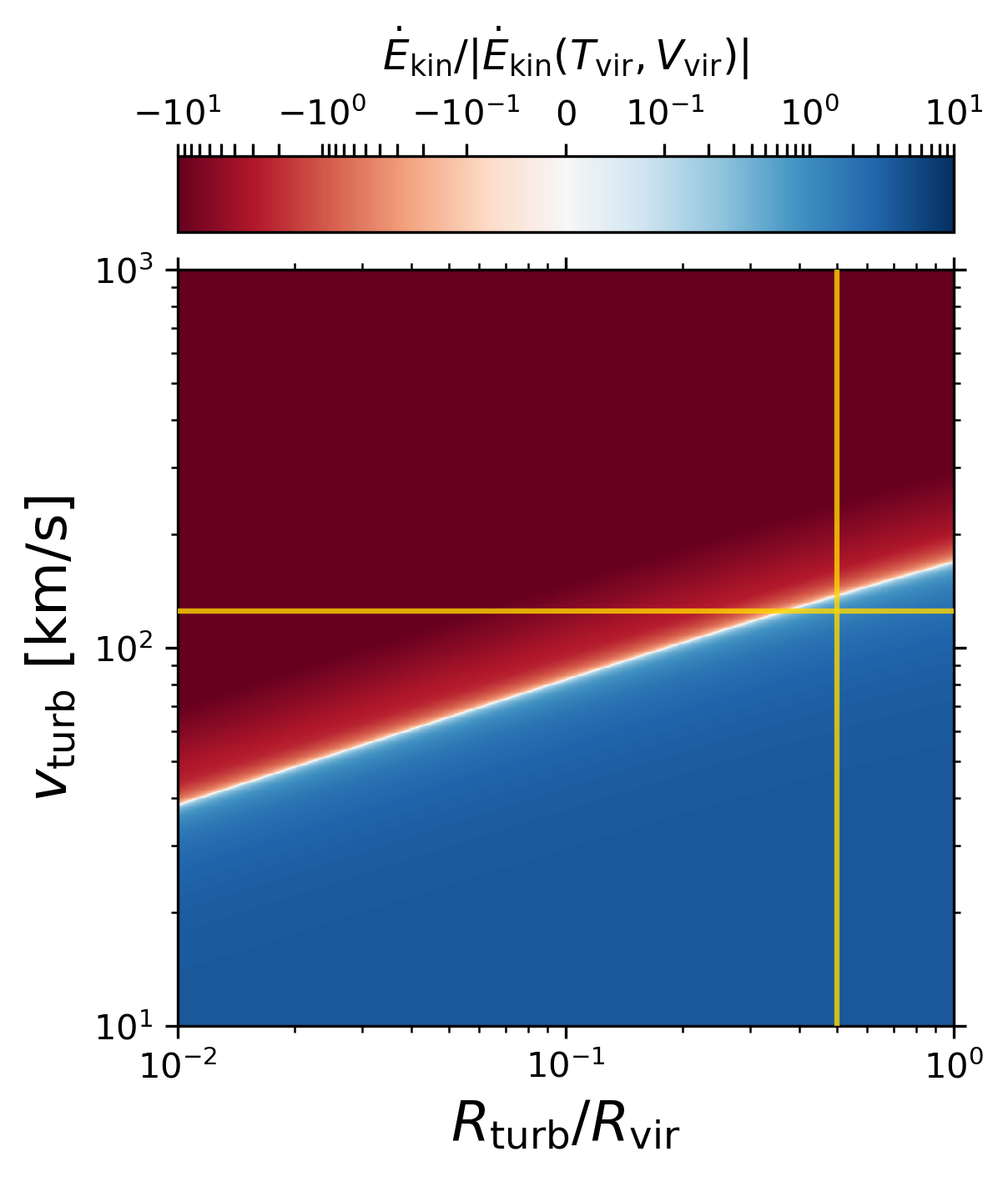}

\caption{Dependence of model equilibria on wind specific energy (top row) and largest eddy turnover scale (bottom row) for the same $z=0$ MW halo parameters as in Figure \ref{fig:heatmap}. The effect of parameter variations on $\dot{E}_{\rm th}$ is shown in the left column and on $\dot{E}_{\rm kin}$ in the right column. The vertical yellow lines mark our fiducial choice for each parameter and the horizontal yellow lines mark $T_{\rm vir}$ and $V_{\rm vir}$. See Section \ref{sec:eq1D} for a description of the effects of each parameter variation.}
\label{fig:eq1D}
\end{figure}

The bottom-left panel of Figure \ref{fig:eq1D} shows how $\dot{E}_{\rm th}$ changes as we simultaneously vary $T_0$ and $R_{\rm turb}$ while fixing $v_{\rm turb}=V_{\rm vir}$. We vary the largest eddy turnover scale $R_{\rm turb}$ between $0.01R_{\rm vir}$ and $R_{\rm vir}$. On the smaller end of assumed values for $R_{\rm turb}$, the eddy turnover time will be faster so there will be more heating from turbulence dissipation and this drives up the equilibrium $T_0$. Dropping $R_{\rm turb}\lesssim0.2R_{\rm vir}$ would lead to $T_0\gg10^6$ K unless we also decrease $v_{\rm turb}$. On the other hand, the equilibrium $T_0$ is roughly flat over the range $R_{\rm turb}\approx0.2-1.0R_{\rm vir}$ likely because $\dot{E}_{\rm diss}\ll\dot{E}_{\rm wind}$ in this regime and thus it is mainly competition between $\dot{E}_{\rm wind}^{\rm th}$ and $\dot{E}_{\rm cool}$ that sets $T_0$ when $R_{\rm turb}$ is large. 

The bottom-right panel of Figure \ref{fig:eq1D} shows how the equilibrium $v_{\rm turb}$ depends on $R_{\rm turb}$. As expected, there is a linear relation between the equilibrium $v_{\rm turb}$ and $R_{\rm turb}$ because these two are directly connected via the eddy turnover time to set the dissipation rate. When $R_{\rm turb}$ is low, turbulence will decay faster and since we have fixed all other parameters that govern turbulence driving, the equilibrium $v_{\rm turb}$ will be lower (and vice versa for larger $R_{\rm turb}$).

\section{Calibrating the model with FIRE}\label{sec:fire}
Now that we have a sense of how the model scales, we turn to setting the unknown free parameters. We could use observations to calibrate some of our model parameters \citep{carr23} or use scaling relations of bulk galaxy/halo properties from large-volume cosmological simulations. However here we take a somewhat different approach and extract the time evolution of various properties for individual galaxies in the FIRE-2 suite. We will use the time evolution of the individual FIRE-2 halos to then set the parameters of our ODEs. We emphasize that the primary goal is to demonstrate the expressiveness of the physical model and not simply to reproduce the FIRE-2 simulations.

\subsection{FIRE-2 Simulations}
We use the ``core'' suite of the FIRE-2 cosmological hydroynamical ``zoom-in'' simulations \citep{hopkins18}. Our sample includes three ultrafaint dwarfs (m10q, m10y, m10z which have $M_{\rm vir}\sim10^{10}M_{\odot}$ at $z=0$), six intermediate-mass dwarfs (m11a, m11b, m11c, m11q, m11v, m11f which have $M_{\rm vir}\sim10^{11}M_{\odot}$ at $z=0$), and three Milky Way-mass halos (m12i, m12f and m12m which have $M_{\rm vir}\sim10^{12}M_{\odot}$ at $z=0$). This is the same sample that we used in \citet{pandya20} and \citet{pandya21} except we exclude the anomalously late-forming low-mass dwarf m10v which has effectively no star formation until $z\lesssim0.5$ (see section 2 of those papers for more details and references about the simulations). We use the same Rockstar halo catalogs and consistent-trees merger trees \citep{behroozi13a,behroozi13b} that we generated and described in \citet{pandya20}. We adopt the \citet{bryan98} definition of virial overdensity.

\subsection{Measuring baryonic properties}
As in \citet{pandya20}, we split each FIRE-2 halo into two zones: gas and star particles within $0.1R_{\rm vir}$ are used to compute the stellar and ISM mass and metallicity of the central galaxy in each snapshot whereas the CGM mass, thermal energy and kinetic energy are computed using all gas particles between $0.1-1.0R_{\rm vir}$. We do not attempt to exclude gas associated with satellites orbiting in the CGM: both their cold gas and hot outflows can affect the global thermodynamics of the host CGM. The global CGM thermal energy is computed by summing over the thermal energy of the individual gas particles:
\begin{equation}
E_{\rm CGM}^{\rm th} = \sum m_i \frac{3}{2} \frac{k_B T_i}{\mu m_p}
\end{equation}
where the subscript $i$ runs over all CGM gas particles, and $m_i$ and $T_i$ are respectively the mass and temperature of the $i$-th particle. Similarly the global CGM kinetic energy is computed as 
\begin{equation}
E_{\rm CGM}^{\rm kin} = \sum m_i \frac{1}{2} v_i^2
\end{equation}
where $v_i^2$ is the norm of the halo-centric particle velocity vector. Note that we do not attempt to decompose the CGM velocity field into a turbulent component so our measurement of $E_{\rm CGM}^{\rm kin}$ includes contributions from bulk and rotating flows. Hence we are measuring an upper limit to the actual turbulent kinetic energy that we care about for comparison to our model. Finally, the stellar, ISM and CGM metal masses are computed very similarly but summing over only the metal mass fractions of the relevant particles. The metallicities are then computed as the ratio of the total zone metal mass divided by the total zone mass further normalized to solar metallicity $Z_{\odot}=0.02$.

We also compute gas inflow and outflow rates at the galaxy and halo scale following \citet{pandya20} and \citet{pandya21}. Complementing those earlier papers, we have implemented single-adjacent-snapshot particle tracking which we will extend in the future to operate over multiple snapshots for the purpose of characterizing halo outflow recycling fractions and timescales \citep[adapting the methodology of][]{anglesalcazar17,hafen19}. After classifying particles into different zones as described above for any two adjacent snapshots, we compute the intersection of the particle ID arrays to identify which particles crossed zones. For example, inflowing particles at the halo scale are those that crossed from $>R_{\rm vir}$ in the first snapshot to $<R_{\rm vir}$ in the later snapshot. We can then estimate the mass inflow rate into the CGM by adding up the masses of all the crossing particles and dividing by the timestep between these two adjacent snapshots \citep[see also][]{wright20}. We repeat this procedure to compute the mass outflow rate from the CGM at $R_{\rm vir}$ and the mass inflow and outflow rates at $0.1R_{\rm vir}$. The gross DM accretion rate at $R_{\rm vir}$ is also computed in this way.

We additionally compute the metal mass flow rates by summing only over the metal mass fractions of the crossing particles at both interfaces. We also compute the energy transported by galactic winds at $0.1R_{\rm vir}$ and at $R_{\rm vir}$ following \citet{pandya21} by multiplying the mass of each crossing particle by its Bernoulli velocity squared, which is a measure of the energy available to drive outflows. We also compute the energy inflow rate at the galaxy and halo scale using the same $v_B$ definition but for particles that are crossing inwards. Finally, since $v_{\rm B}^2$ is the sum of the specific kinetic energy flux and the specific enthalpy flux, we can compute the fraction of the energy flux that is already thermalized by taking the ratio of the specific enthalpy flux to the Bernoulli velocity squared ($1.5c_s^2 / v_{\rm B}^2$ following equation \ref{eqn:vB}). This constrains $f_{\rm thermal}^{\rm wind}$ and $f_{\rm thermal}^{\rm accretion}$ for our model. 

Our results are very similar to the mass fluxes reported in \citet{pandya20} and \citet{pandya21} with the caveat that here we do not impose any cut on $v_B$ for outflowing particles, unlike \citet{pandya21}. Thus our outflow rates correspond to simply using a $v_{\rm rad}>0$ km/s threshold. This is done to ensure that when we assume the FIRE-2 wind loading factors in our model, we will conserve mass during the gas flow cycle. Otherwise, a significant cut on $v_B$ for winds as in \citet{pandya21} may miss up to half of the outflowing gas going into the inner CGM and this would make it impossible for our model to reproduce the time evolution of the CGM and ISM mass measured in the simulations. As a consistency check, we have verified that integrating our measured mass fluxes gives a nearly identical CGM and ISM mass as a function of time as measured from the particle data. This exercise increases our confidence in our fluxes and bulk measurements in the sense that if our model can accurately reproduce the fluxes as a function of time, then it should naturally also reproduce the time evolution of the integrated CGM and ISM properties.

\subsection{Model parameters}

\subsubsection{Star formation and SN-driven winds}
\begin{figure*}
\centering
\includegraphics[width=\hsize]{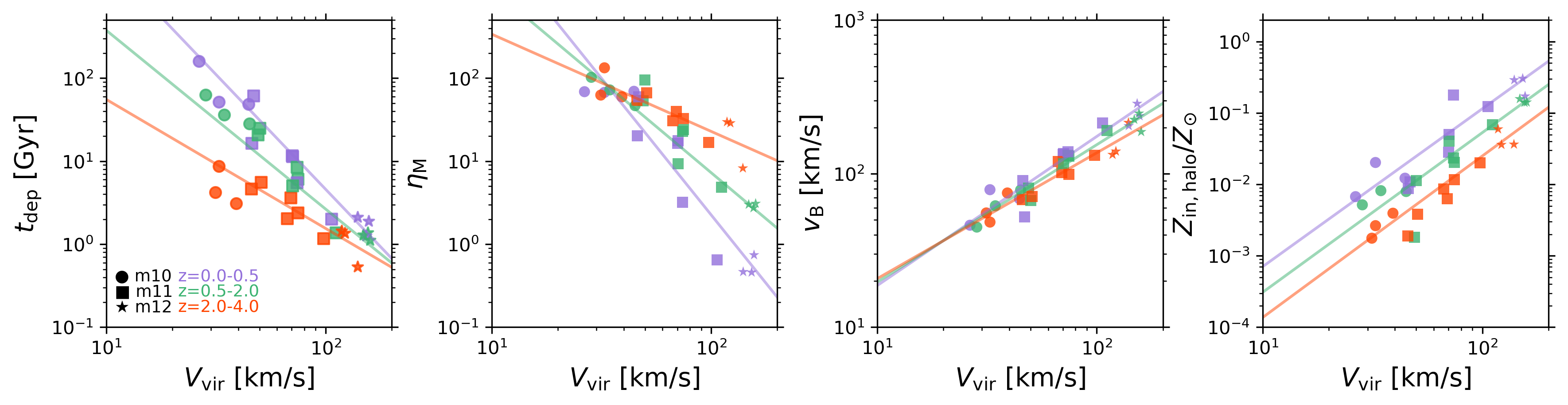}
\caption{Functional forms of four key model parameters derived from FIRE-2. Left: ISM depletion time vs. $V_{\rm vir}$ and redshift from FIRE-2. Second from left: wind mass loading factors parameterized from FIRE-2. Third from left: Bernoulli velocity of winds leaving the ISM in FIRE-2. Right: metallicity of inflowing gas at the halo virial radius in FIRE-2.}
\label{fig:params_FIRE}
\end{figure*}

In our model, the small-scale physics of SF and SN-driven winds is represented by three parameters: the depletion time of the entire ISM mass ($t_{\rm dep}$), the wind mass loading factor ($\eta_{\rm M}$), and the wind specific energy as quantified by the Bernoulli velocity ($v_B$). Figure \ref{fig:params_FIRE} shows our fit to these parameters using the FIRE-2 data. We parameterize $t_{\rm dep}$ as
\begin{equation}\small
\frac{t_{\rm dep}}{\rm{Gyr}} = 10^{0.46} \left(\frac{V_{\rm vir}}{125\,\rm{km/s}}\right)^{-3.0 + 2.4\log(1+z)} (1+z)^{-0.7}
\end{equation}
We parameterize $\eta_{\rm M}$ as 
\begin{equation}
\eta_{\rm M} = 10^{-0.2} \left(\frac{V_{\rm vir}}{125\,\rm{km/s}}\right)^{-3.7 + 4.2\log(1+z)} (1+z)^{2.4}
\end{equation}
and the Bernoulli velocity as 
\begin{equation}\small
v_{\rm B} = 10^{2.36}\left(\frac{V_{\rm vir}}{125\,\rm{km/s}}\right)^{1.0 - 0.3\log(1+z)}(1+z)^{-0.24}
\end{equation}
Although we could equivalently parameterize $\eta_{\rm E}$ instead of $v_B$, we find that $v_B$ can more easily be fit with a simple power law whereas $\eta_{\rm E}$ requires a more complicated fit \citep[see also Figure 9 of][]{pandya21}.

\subsubsection{Chemical Evolution}
Recall that our model has two free parameters governing chemical evolution: (1) the nucleosynthetic yield which we take to be $(1-f_{\rm rec})y_Z=0.02$ as appropriate for a \citet{kroupa01} IMF, and (2) the metallicity of gas flowing into the halo. In principle we could also introduce another free parameter for the wind metal enrichment factor but we avoid that complication and assume winds have the same metallicity as the ISM. This may be particularly justified for FIRE-2 for which we found that the winds were so heavily mass-loaded that their average metallicity probably tracks that of the entrained ISM \citep{pandya21}. Figure \ref{fig:params_FIRE} shows how $Z_{\rm in,halo}$ scales with $V_{\rm vir}$ and redshift for the FIRE-2 halos. We find that 
\begin{equation}\small
\frac{Z_{\rm in,halo}}{Z_{\odot}} = 10^{-0.6} \left(\frac{V_{\rm vir}}{125\,\rm{km/s}}\right)^{2.2 + 0.1\log(1+z)} (1+z)^{-1.3}
\end{equation}
The halo inflow metallicities in FIRE-2 are non-zero and depend strongly on both $V_{\rm vir}$ and redshift. The non-zero metallicities likely reflect both recycling of galactic winds on large-scales as well as chemical pre-enrichment by neighboring halos. The redshift dependence at fixed $V_{\rm vir}$ can be understood as there simply being more metals produced and hence available to be recycled at later times. The $V_{\rm vir}$ dependence likely arises for multiple reasons: (1) wind metallicities are higher in more massive halos as shown in the bottom panel, (2) more massive halos have higher SFRs and hence higher wind mass outflow rates even though their mass loading factors are lower, and (3) more massive halos have more satellites that can pollute their local environment before accretion. Still, it is striking that in FIRE-2, the MW-mass halos at $z\sim0$ have inflowing metallicities at $R_{\rm vir}$ of order $\sim0.1Z_{\odot}$. 

\subsubsection{Preventative feedback for halo gas accretion}
The leftmost panel of Figure \ref{fig:logistics} shows our preventative feedback parameter $f_{\rm prev}$ from equation \ref{eqn:Mdot_in_halo}. The data points are measurements of the ratio $\dot{M}_{\rm in,gas}/(f_b \dot{M}_{\rm in,gross})$ at $R_{\rm vir}$ for individual FIRE-2 halos at $z=0$ and $z=3$. We use a simple generalized logistic function to approximately capture this trend:
\begin{equation}\small\label{eqn:logistic}
f_{\rm prev} = 0.25 + \frac{0.96 - 0.25}{1.0 + \exp[-7.13(\log V_{\rm vir} - 1.88)]}
\end{equation}
In \citet{carr23}, we self-consistently predict $f_{\rm prev}$ using the ratio of $E_{\rm out,halo}$ to $E_{\rm in,halo}$ \citep[see also section 5.2 of][]{pandya20}. In order for that model to capture the complexity of FIRE-2, we would need to introduce a free parameter that quantifies how much of the outflowing energy couples to and heats the accreting gas. This parameter may have a complicated dependence on halo mass, redshift, etc. with scatter. Thus we opt for a simpler parameterization in this work. 

\subsubsection{Thermal vs. kinetic flux partitioning}
The middle two panels of Figure \ref{fig:logistics} parameterize the fraction of cosmic accretion energy and SN wind energy that is thermal for the FIRE-2 halos. We assume
\begin{equation}\small
f_{\rm thermal}^{\rm accretion} = 10^{-0.17} \left(\frac{V_{\rm vir}}{125\,\rm{km/s}}\right)^{\alpha(z)} (1+z)^{-0.66}
\end{equation}
where $\alpha(z)= -0.28 - 0.87\log(1+z)$ captures the redshift dependence of the slope. At low $V_{\rm vir}$, this power law can lead to values larger than one; in this case we simply set $f_{\rm thermal}^{\rm accretion}=1$ to prevent negative $\dot{E}_{\rm in,halo}^{\rm kin}=(1-f_{\rm thermal}^{\rm accretion})\dot{E}_{\rm in,halo}$. The decrease in $f_{\rm thermal}^{\rm accretion}$ with $V_{\rm vir}$ and redshift reflects the importance of cold filamentary accretion in the FIRE-2 halos.

For $f_{\rm thermal}^{\rm wind}$ we use a generalized logistic function:
\begin{equation}\small
f_{\rm thermal}^{\rm wind} = 0.35 + \frac{1.0 - 0.35}{1.0 + \exp[20.88(\log V_{\rm vir} - 1.5)]}
\end{equation}
It is interesting that ultrafaints have thermally-supported winds whereas classical dwarfs and MW-mass galaxies all cluster around $f_{\rm thermal}^{\rm wind}\approx0.35$, indicating efficient mixing and/or cooling. In Appendix \ref{sec:appendix}, we will show how our predictions change if we set $f_{\rm thermal}^{\rm accretion}=1$ and $f_{\rm thermal}^{\rm wind}=1$ so that we are in the purely thermal limit of our model \citep[as in our companion paper,][]{carr23}.

\subsubsection{Characteristic size of largest turbulent eddies}\label{sec:Rturb}
\begin{figure*}
\centering
\includegraphics[width=\hsize]{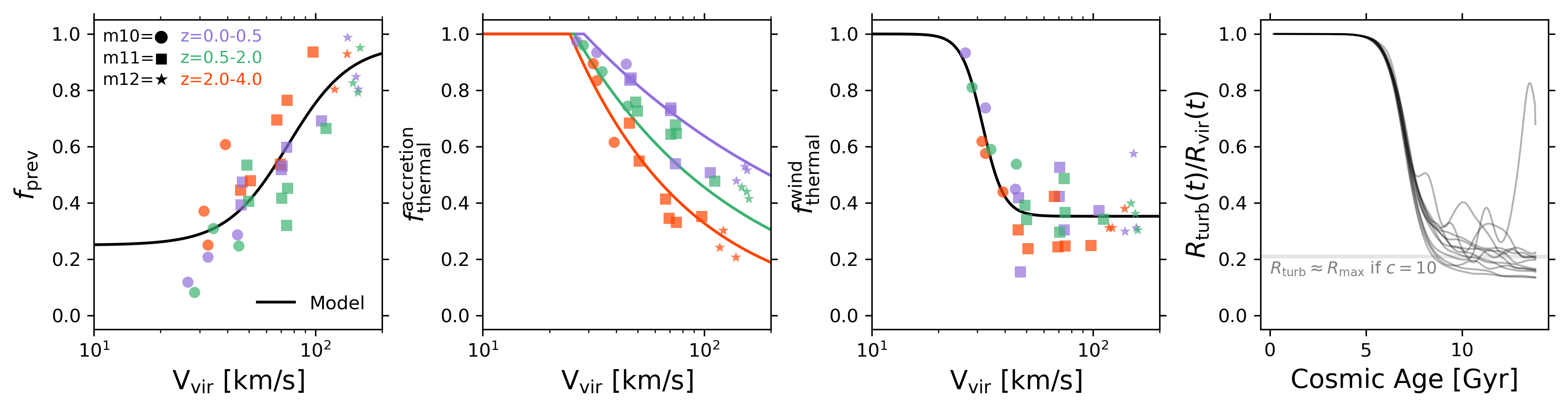}
\caption{Assumed functional forms for the remaining free parameters of our model. Left: preventative feedback parameter to suppress halo gas accretion. The solid line shows our logistic function and the points show the ratio $\dot{M}_{\rm in,gas}/(f_b \dot{M}_{\rm in,gross})$ at $R_{\rm vir}$ for the FIRE-2 halos at $z=0$ (black) and $z=3$ \citep[magenta; see also Figure 14 of][]{pandya20}. Second from left: fraction of cosmic accretion energy that is thermal. Third from left: fraction of SN wind energy that is thermal. Right: size of the largest CGM eddies normalized by halo virial radius as a function of time. We assume $R_{\rm turb}\approx R_{\rm vir}$ at early times which may be expected if CGM turbulence is primarily driven by cosmic accretion but that at late times $R_{\rm turb}\approx R_{\rm max}$ if turbulence is predominantly driven in the inner CGM by SN winds.}
\label{fig:logistics}
\end{figure*}

The characteristic size of the largest turbulent eddies in the CGM is one of the main uncertainties of our model since we do not have existing constraints on this observationally or from cosmological simulations. The rightmost panel of Figure \ref{fig:logistics} shows our assumed generalized logistic function for $R_{\rm turb}(t)$ normalized by $R_{\rm vir}(t)$: 
\begin{equation}
R_{\rm turb}(t) = R_{\rm max} + \frac{R_{\rm vir} - R_{\rm max}}{1.0 + \exp(2(t - 7.0\,\rm{Gyr}))}
\end{equation}
where again $R_{\rm max}$ is the radius of $V_{\rm max}$ as used in equation \ref{eqn:tffeff}. The slope (2) and pivot (7 Gyr) of our logistic function were arbitrarily chosen and control when and how quickly $R_{\rm turb}$ drops from $R_{\rm vir}\to R_{\rm max}$. Note that if we replace $R_{\rm max}$ with $R_{\rm vir}$, then we would simply have $R_{\rm turb}(t)=R_{\rm vir}(t)$. We imagine that at early times, large-scale structure formation continuously drives turbulence such that $\dot{E}_{\rm diss}\approx0$ which is achieved as $R_{\rm turb}\to\infty$. On the other hand, at later times, we assume that SN winds are the primary drivers of turbulence and that this occurs mainly in the inner halo such that $R_{\rm turb}\approx R_{\rm max}$. Our model is quite sensitive to the exact choice of these asymptotic values -- smaller $R_{\rm turb}$ at early times leads to faster turbulence dissipation and thus less turbulent pressure support, higher accretion rates into the ISM, and more star formation.

\section{Reproducing the time evolution of individual simulated FIRE-2 halos}
As a first application, we explore how well our FIRE-2 calibrated model compares to the actual evolution of the simulated FIRE-2 halos. Our goal in this section is to show that the model follows the general trends of the simulations to within a factor of a few but also to identify discrepancies which can guide future improvements to the model.

\subsection{Mass assembly histories}
\begin{figure*}
\centering
\includegraphics[width=0.99\hsize]{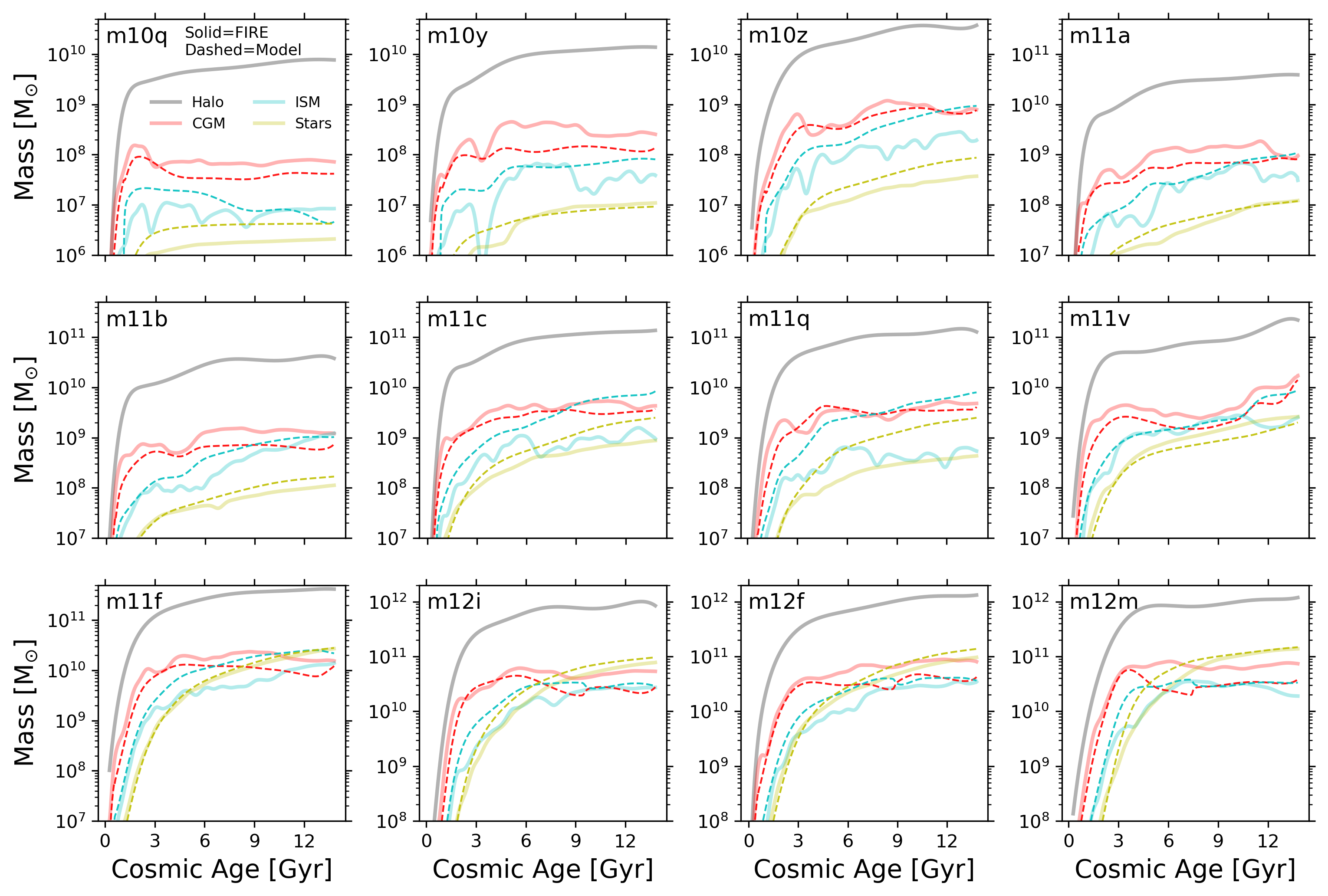}
\caption{Time evolution of DM, CGM, ISM and stellar mass for the individual FIRE-2 halos. Actual FIRE-2 measurements are shown as the solid lines and the model predictions are the dashed lines. We see good agreement with FIRE-2 (to within a factor of a few) as a function of time.}
\label{fig:masses}
\end{figure*}

Figure \ref{fig:masses} shows the time evolution of DM, CGM, ISM and stellar mass for the 12 individual core FIRE-2 halos as measured from the particle data and as predicted by our model. The model trajectories generally follow the time series of mass measurements from the FIRE-2 particle data to within a factor of a few. However, there are exceptions: the CGM mass tends to be systematically underestimated by a factor of two in the model relative to FIRE-2, and the stellar and ISM masses of some dwarfs are larger than FIRE-2 by a factor of up to $\sim10$. Some of these discrepancies may be due to our simple average parameter fits that do not capture scatter between individual halos (Figures \ref{fig:params_FIRE} and \ref{fig:logistics}). More importantly, for parameters that could not easily be measured from the simulations such as the turbulence dissipation timescale or an effective free-fall radius for computing the ISM accretion rate, our choices may not reflect what is happening in the simulations. By comparing the flow rates of mass, energy and metals between the model and FIRE-2 in the next few subsections, we can learn more about the behavior of the new model and possible causes for the discrepancies in the mass assembly histories.

\subsection{Halo baryon fractions}
\begin{figure}
\centering
\includegraphics[width=0.99\hsize]{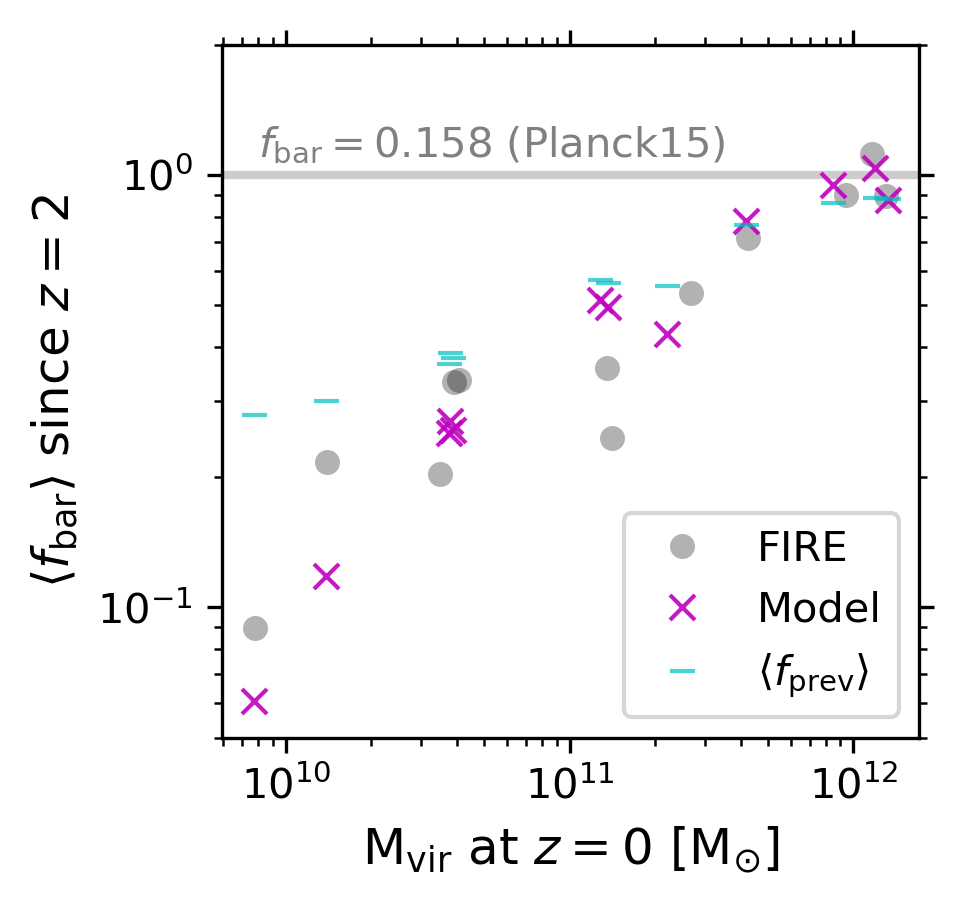}
\caption{Average baryon fraction since $z=2$ as a function of $z=0$ virial mass for the FIRE-2 halos as measured from the particle data (gray circles) and as predicted by our model (magenta crosses). Both FIRE-2 and our model predict that dwarfs have reduced baryon fractions. In our model, this happens because of two mechanisms: (1) preventative feedback suppresses cosmic accretion in dwarfs, and (2) the CGM can become overpressurized so previously accreted baryons are ejected. The cyan lines show the average $f_{\rm prev}$ since $z=2$ which provides a rough upper limit on the halo baryon fraction; the lowest mass halos have $f_{\rm b} < f_{\rm prev}$ which denotes the importance of the CGM overpressurization channel.}
\label{fig:fbar}
\end{figure}

Since we roughly reproduce the time evolution of the masses of the different components of the FIRE-2 halos, it is natural to ask what our model implies for the baryon fractions of halos. Figure \ref{fig:fbar} shows the average halo baryon fraction since $z=2$ as measured from the particle data and as predicted by our model. We have verified that the baryon fractions are roughly constant since $z=2$ so taking the average is a good summary statistic. We reproduce the trend in FIRE-2 which is that dwarfs have $\lesssim10\%$ of the universal baryon fraction $f_{\rm b}=0.158$. In our model, this happens because of two things: (1) the preventative feedback parameter $f_{\rm prev}$ suppresses cosmic accretion, and (2) the CGM can become overpressurized which ejects previously accreted baryons. Since $f_{\rm prev}$ provides a rough upper limit on the halo baryon fraction, having $f_{\rm b}<f_{\rm prev}$ indicates that CGM overpressurization is important, and this is indeed the case for the lowest mass halos in our model.

\subsection{Mass flow rates}\label{sec:mdots}
\begin{figure*}
\centering
\includegraphics[width=0.99\hsize]{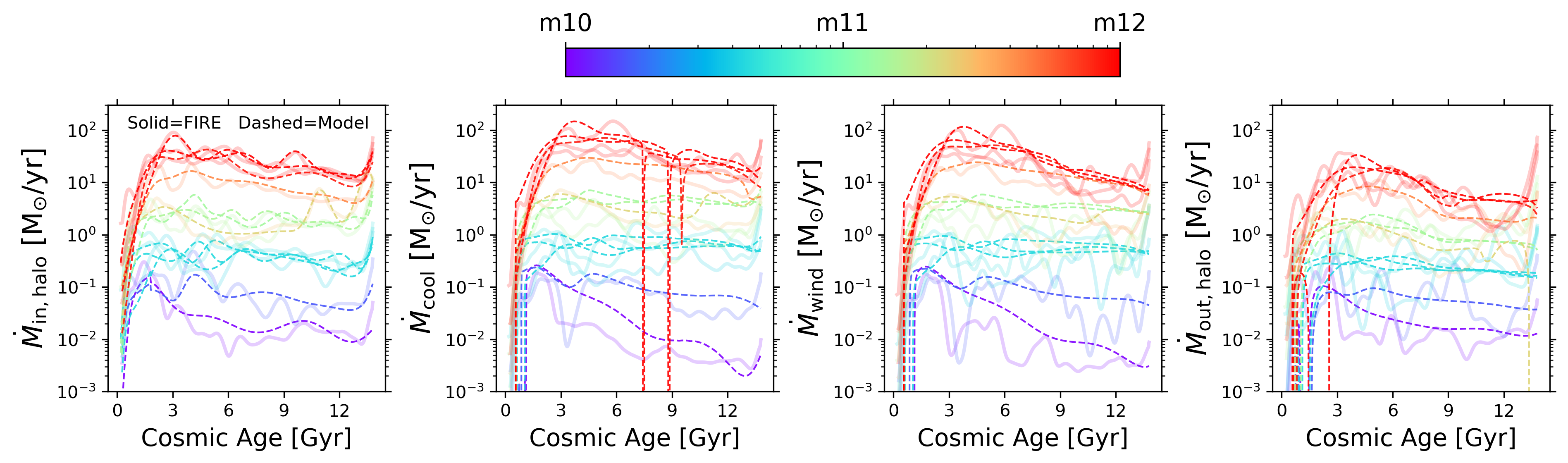}
\caption{Time evolution of the four mass fluxes that underlie the $\dot{M}_{\rm CGM}$ and $\dot{M}_{\rm ISM}$ time derivatives for the individual FIRE-2 halos. The actual measurements from the particle data are solid lines and the model predictions are the dashed lines. The cooling rate and halo outflow rate are genuine predictions of our CGM model and match FIRE-2 very well. On the other hand, it is not surprising that our halo accretion rate and ISM winds match FIRE-2 since we calibrated $f_{\rm prev}$ and $\eta_{\rm M}$ from FIRE-2. Note that the dip in $\dot{M}_{\rm cool}$ for the MW-mass halos happens when $\dot{E}_{\rm cool}<\dot{E}_{\rm diss}$ thus causing $t_{\rm cool,eff}\to\infty$ in Equation \ref{eqn:tcooleff}.}
\label{fig:mdots}
\end{figure*}

Matching the evolution of bulk masses in itself is not sufficient to claim that our model reproduces FIRE-2 since there are many different ways to get to the same mass \citep{pandya20}. Figure \ref{fig:mdots} shows that we also roughly reproduce the time evolution of the individual mass flow rates underlying the $\dot{M}_{\rm CGM}$ and $\dot{M}_{\rm ISM}$ time derivatives. Since we parameterized $f_{\rm prev}$ and $\eta_{\rm M}$ from FIRE-2, it is perhaps not surprising that we reproduce the cosmic halo accretion rate and ISM wind mass loss rate. However, the cooling rate and halo outflow rate are genuine predictions of our model and match FIRE-2 quite well. The main exception is that our cooling rates tend to be on the higher side for the dwarfs but these are sensitive to our predictions for the turbulence dissipation rate and choice of an effective free-fall radius for computing $\dot{M}_{\rm cool}$. It is possible to achieve better agreement with FIRE-2 by increasing $R_{\rm turb}$ or $r_{\rm ff}$ since the former would provide more turbulent pressure support (because the turbulence would decay more slowly) and the latter would increase the effective free-fall time of CGM gas, in turn lowering $\dot{M}_{\rm cool}$, SFR, $\dot{M}_{\rm wind}$ and probably also $\dot{M}_{\rm out,halo}$. But even with our fiducial choices, the results of the simple model show promise for reproducing mass flows in FIRE-2.

\subsection{Chemical evolution}
Having shown that our model is capable of roughly reproducing the mass budgets and mass flow rates of FIRE-2, we now turn to the metal budgets and metal flow rates. Metals are an important additional dimension to predict because they will unlock more observables that we can use to eventually test our model.

\begin{figure*}
\centering
\includegraphics[width=0.99\hsize]{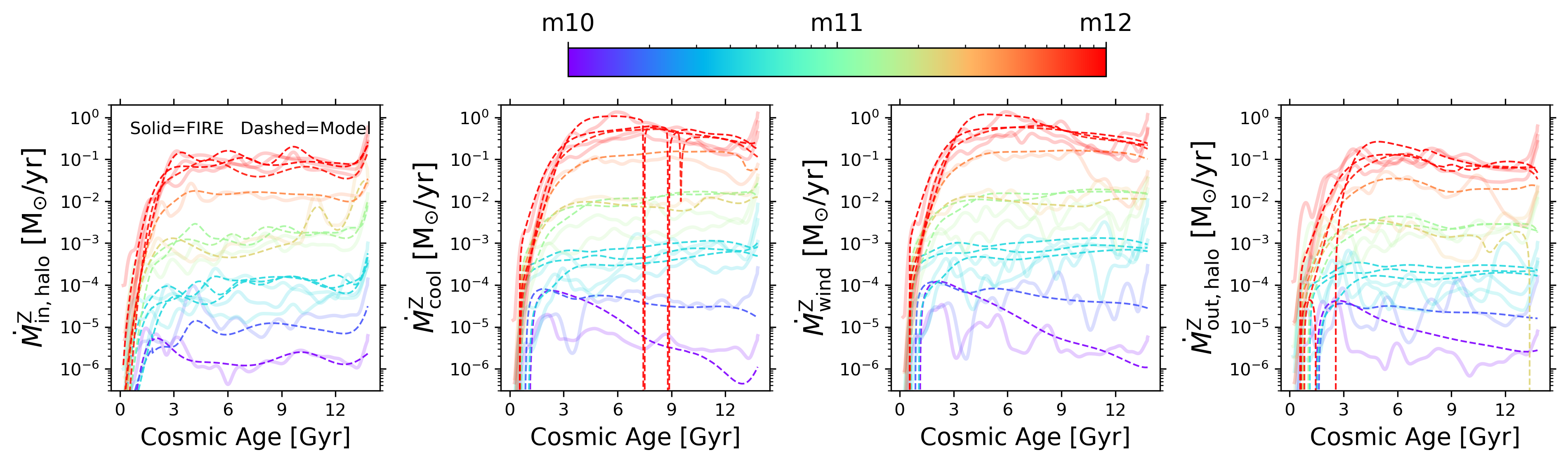}
\caption{Similar to Figure \ref{fig:mdots} but now for the metal mass flow rates. Our model agrees very well with FIRE-2.}
\label{fig:mzdots}
\end{figure*}

\begin{figure*}
\centering
\includegraphics[width=0.99\hsize]{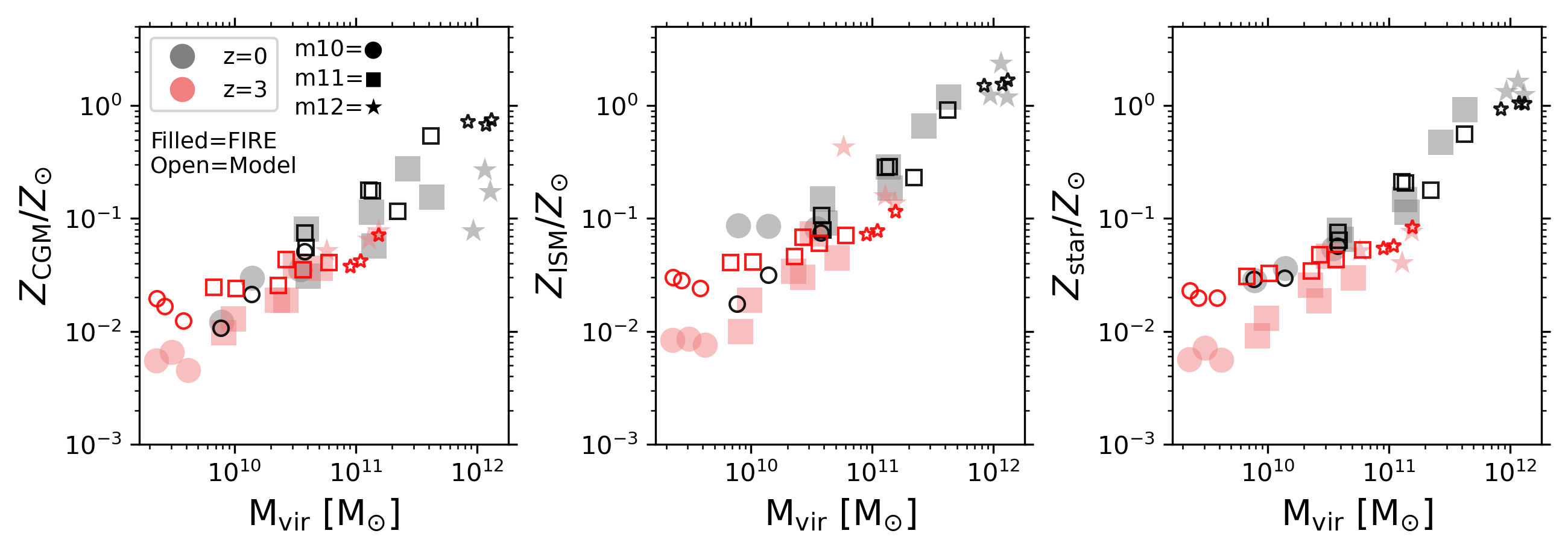}
\caption{Comparing the bulk metallicity of the CGM (left), ISM (middle) and stars (right) as a function of halo mass between our model and FIRE-2. Black is $z=0$ and red is $z=3$ whereas filled symbols show FIRE-2 and open symbols show our model. The CGM metallicities agree well at both low and high redshift. The ISM and stellar metallicities also roughly agree which noteworthy given the simplicity of our chemical evolution model. Variations in the wind enrichment factor, uncertainties in the nucleosynthetic yield parameter, and a breakdown of the instantaneous recycling approximation particularly at high-redshift may help explain remaining discrepancies.}
\label{fig:ZMvir}
\end{figure*}

Figure \ref{fig:mzdots} compares the time evolution of metal flow rates from our model to FIRE-2. We see very good agreement, as with the overall mass flow rates. Figure \ref{fig:ZMvir} plots the bulk metallicities of our three components (CGM, ISM and stars) as a function of halo mass in comparison to FIRE-2. We show the comparison for $z=0$ and a representative high redshift $z=3$. The model tracks the general trends of CGM, ISM and stellar metallicity with halo mass at both low and high redshift, which is noteworthy given the simplicity of our chemical evolution prescriptions. One discrepancy is that the CGM metallicity of the model MW halos is $\sim4\times$ higher than FIRE-2 whereas the ISM and stellar metallicities agree to better than a factor of two. The high CGM metallicity of the MW halos at $z=0$ partially reflects the CGM mass being a factor of $\sim2$ lower in the model compared to FIRE-2 (see Figure \ref{fig:masses}) with the remaining excess possibly attributable to our assumption that the wind metallicity equals the ISM metallicity (it may be lower if there was entrainment of inner CGM gas or if the ISM is inhomogeneous and winds are launched from less-enriched regions). The other main discrepancy is that the lowest mass halos at $z=3$ have higher metallicities in the model compared to FIRE-2. This may reflect a breakdown of the instantaneous recycling approximation at high redshift.

\subsection{CGM energetics and phase transitions}
\begin{figure*}
\centering
\includegraphics[width=0.99\hsize]{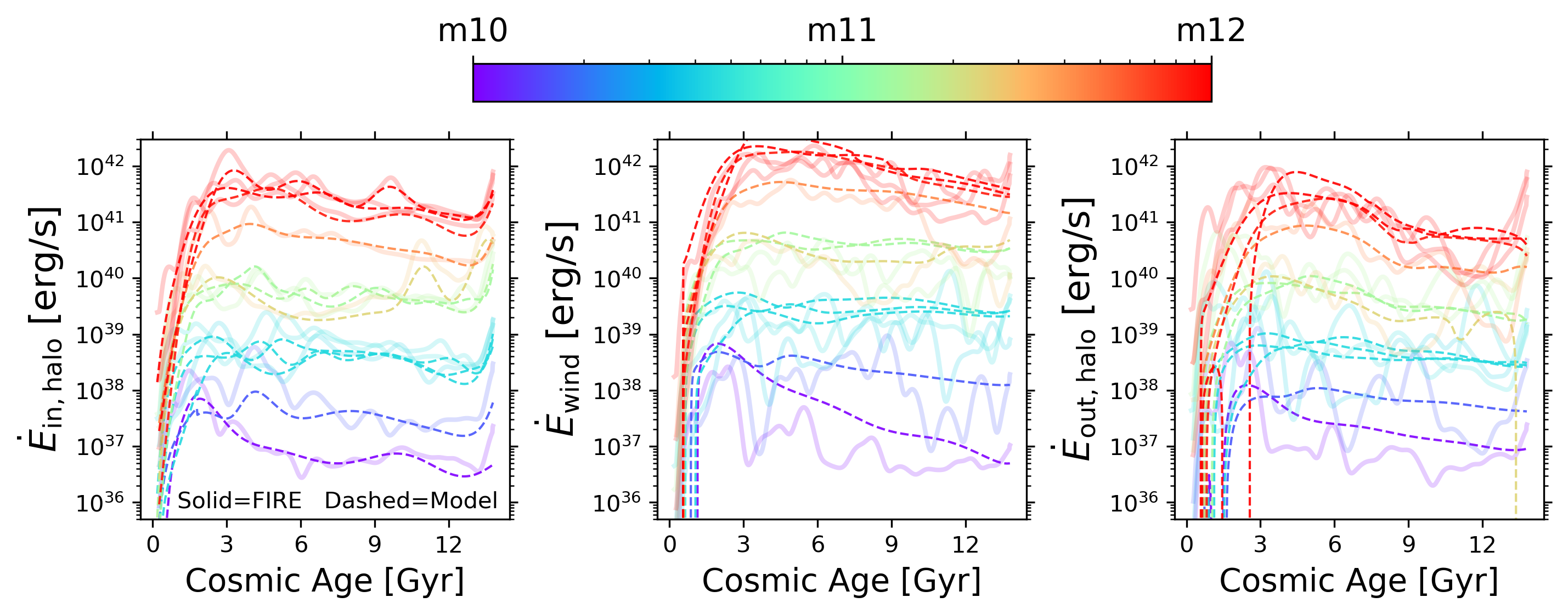}
\caption{Time evolution of the energy inflow rate from cosmic accretion (left), SN-driven winds (middle) and halo-scale energy outflow rate due to CGM overpressurization (right). As in previous figures, the solid lines are measurements from the FIRE-2 simulations whereas the dashed lines are the predictions from our model. Overall, the model agrees quite well with the simulations over a large range in halo mass and time.}
\label{fig:Edots}
\end{figure*}

Figure \ref{fig:Edots} shows the time evolution of the energy inflow rate from cosmic accretion and SN-driven winds as well as the halo-scale energy outflow rate due to CGM overpressurization. Here again the inflow rate of energy from cosmic accretion agrees to better than a factor of two between the model and FIRE-2 whereas the wind and halo energy outflow rates show some discrepancies. First, while the wind energy outflow rates for most halos agree with FIRE-2 to within a factor of a few, the lowest mass dwarf shows an excess in the model. This directly follows from its excess CGM cooling rate and excess SFR which could be resolved by increasing the turbulence dissipation time and/or effective free-fall time to limit the ISM accretion rate as discussed in subsection \ref{sec:mdots}. Second, the halo energy outflow rates also agree with the FIRE-2 measurements to within a factor of a few except at very early times where the model tends to have no halo outflows. This lack of early halo outflows in the model may be due to additional contributions from other sources that are not included (e.g., more turbulence driven by satellite motions). Despite these differences, it is encouraging that the model follows the general trends of the simulations.

\begin{figure*}
\centering
\includegraphics[width=0.99\hsize]{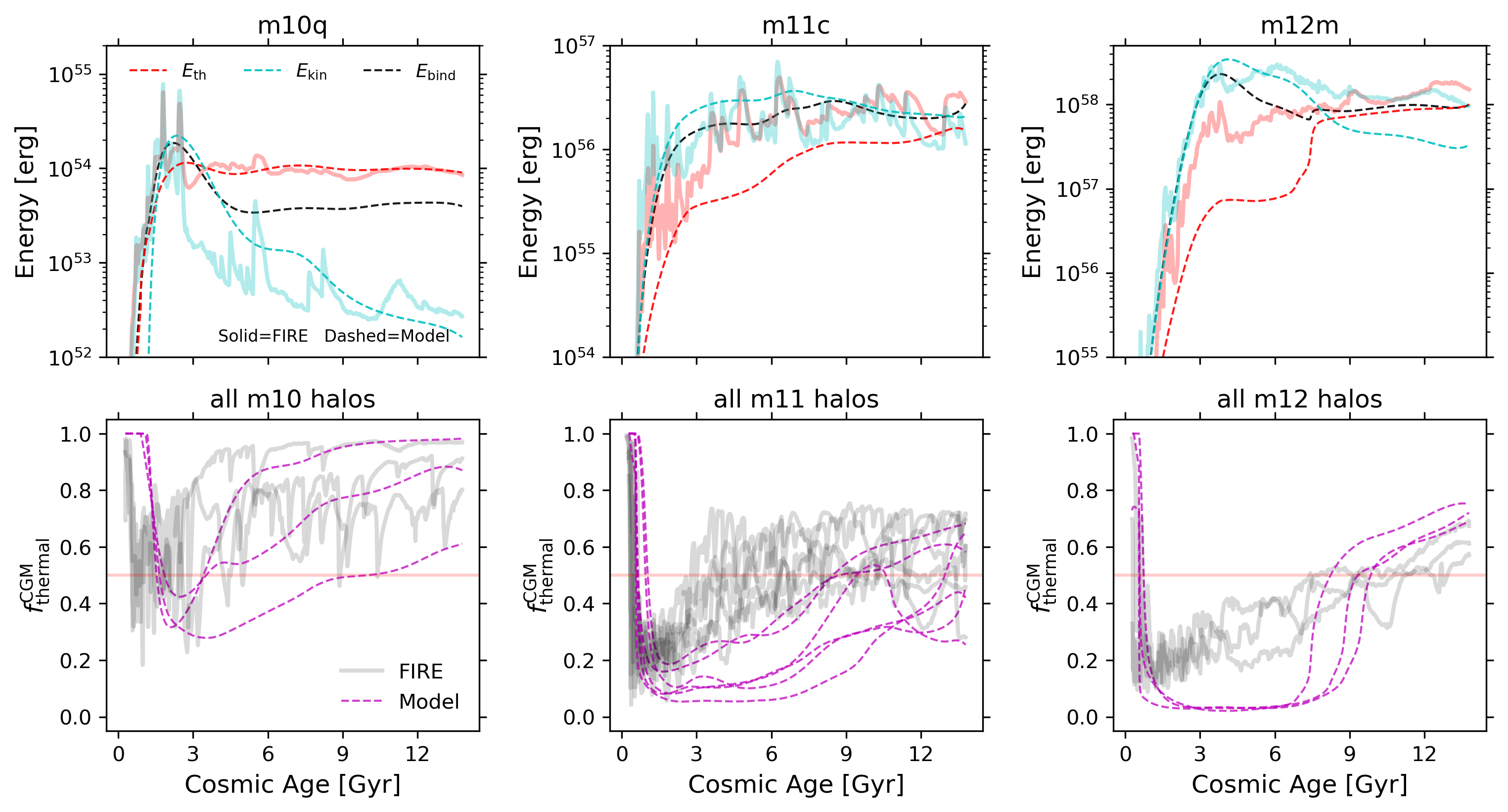}
\caption{Time evolution of CGM thermal energy, kinetic energy and binding energy for three example FIRE-2 halos (top row). The solid lines show our measurements from the particle data and the dashed lines are the predictions of our model. Our model roughly reproduces the trends in the simulations, namely that there is an early predominantly kinetic phase and then a transition to a thermally-supported CGM at some later time. In the low-mass dwarfs, this transition happens quite early so their CGM is purely thermal for most of cosmic time. However in intermediate-mass dwarfs and MW-mass halos, the transition happens at a later time. The bottom row shows the fraction of the global CGM energy in thermal form as a function of time as predicted by our model (dashed magenta lines) and as measured from the particle data (solid gray lines) for all 12 core FIRE-2 halos. The horizontal red line denotes $f_{\rm thermal}^{\rm CGM}=0.5$. The low-mass dwarfs are consistently high, the MW-mass halos show an abrupt transition at $t\sim9$ Gyr ($z\approx0.5$) in the model but rise more gradually in FIRE-2, and the intermediate-mass dwarfs cluster around $f_{\rm thermal}^{\rm CGM}\approx0.5$ in FIRE-2 and show a late, gradual transition in the model.}
\label{fig:energetics}
\end{figure*}

In analogy, Figure \ref{fig:energetics} shows the time evolution of CGM thermal energy, kinetic energy and binding energy both as predicted by our model and as measured in the FIRE-2 particle data for three representative halos. Our model roughly reproduces the trends measured from the FIRE-2 data. All three halos show a transition from an early, cool, kinetic-dominated phase to a thermally-supported warm/hot halo at later times. In the low-mass dwarf, this transition happens very early since cooling can only balance heating from UVB photoionization, SN winds, turbulence dissipation and cosmic accretion at a slightly super-virial temperature. The intermediate-mass dwarf transitions more gradually and at lower redshift since it is at the peak of the cooling curve and hence can offset more of the SN/turbulent heating. The MW-mass halo transitions abruptly at an intermediate redshift in our model but more gradually in FIRE-2. As shown by \citet{stern21}, the transition in FIRE-2 is radially-dependent with the outer CGM undergoing the transition first. This complexity is not reflected in our model which treats the entire CGM as a single zone (thus leading to a sharper transition) and is also not captured by our measurements of a single global $f_{\rm thermal}^{\rm CGM}$ for the FIRE-2 halos in Figure \ref{fig:energetics} (we will discuss this further in subsections \ref{sec:implications} and \ref{sec:limitations}). Nevertheless, our model does capture the essential idea that the CGM may undergo a ``thermalization'' process.

\begin{figure}
\centering
\includegraphics[width=0.99\hsize]{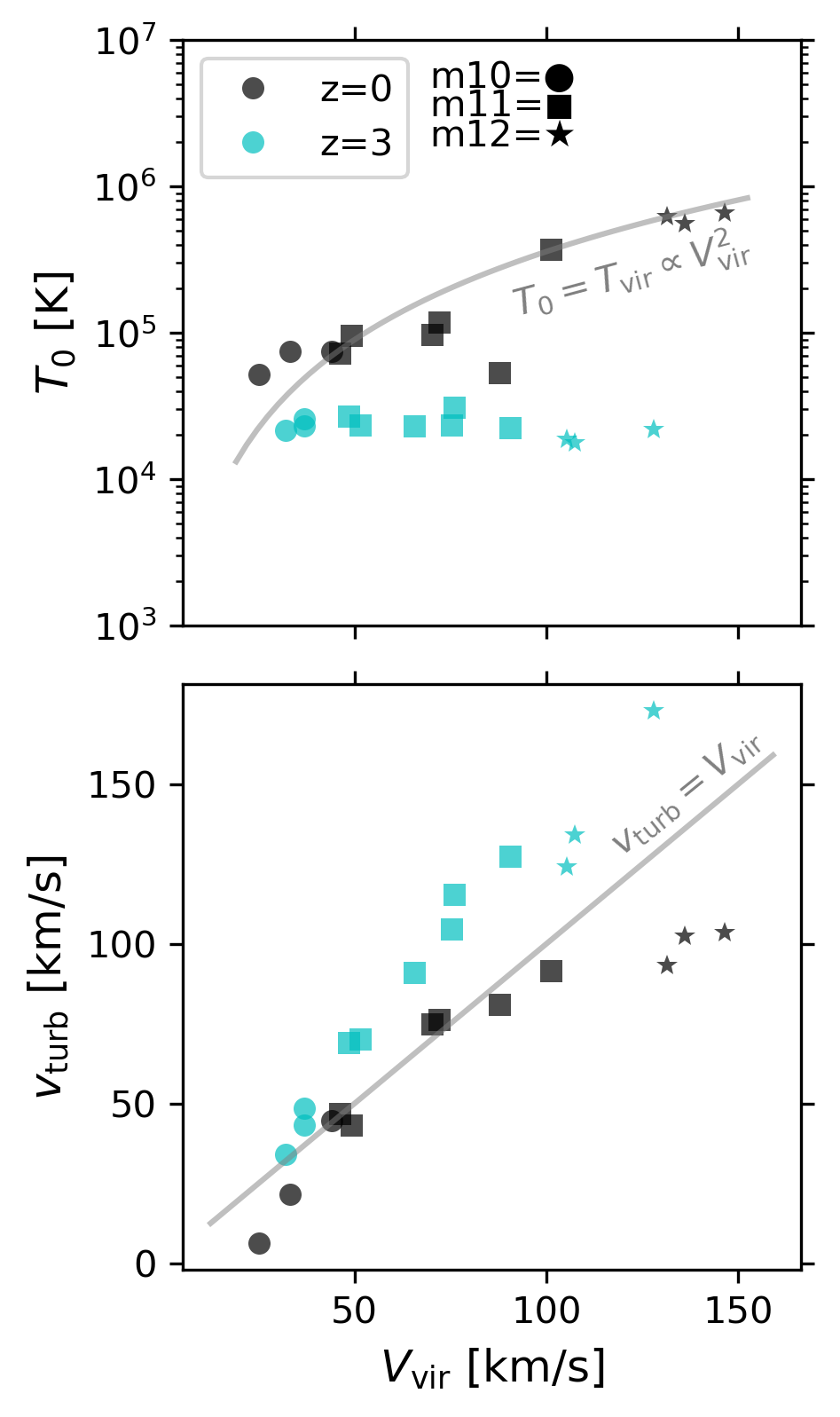}
\caption{The average global CGM temperature (top) and turbulent velocity (bottom) predicted by our model as a function of the halo virial velocity at $z=0$ (black) and $z=3$ (cyan). We can see that at late times, the temperature roughly follows the virial temperature of the halo as expected, but at earlier times the two are decoupled with $T_0\ll T_{\rm vir}$. As for the specific turbulent kinetic energy, $v_{\rm turb}\gtrsim V_{\rm vir}$ for dwarfs at both low and high redshift as well as for MW progenitors. But for MW-mass halos at $z\sim0$, the turbulence has decayed to become only a fraction of the halo specific energy.}
\label{fig:T0vturbVvir}
\end{figure}

Figure \ref{fig:T0vturbVvir} compares the average global temperature and turbulent velocity of the CGM to the halo virial velocity for our halos at both low and high redshift. At late times, the CGM temperature roughly tracks the virial temperature of the dark matter halo as expected, but at early times the two are decoupled with $T_0\ll T_{\rm vir}$. The turbulent velocity is comparable to the halo virial velocity, or even exceeds it, in the dwarfs at both low and high redshift. In the case of the MW-mass halos, this is only true for their progenitors: by late times the turbulent specific energy decays to become only a fraction of the halo specific energy.

\subsection{Time evolution of model equilibria}\label{sec:eqevol}
We saw in the previous subsection that the CGM in our model transitions from an early, cool, turbulent phase to a warm/hot thermally-supported phase at late times. Here we analyze the time evolution of our model equilibria to provide insight on why this CGM phase transition happens. This builds on section \ref{sec:eq} where we showed how to analyze the equilibria of our model using idealized state variables and parameters for a $z=0$ MW-mass halo. The difference is that here the state variables such as CGM mass and metallicity are self-consistently evolved in time and we have also determined our model parameters from FIRE-2 or simple physical arguments.

\begin{figure*}
\centering
\includegraphics[width=\hsize]{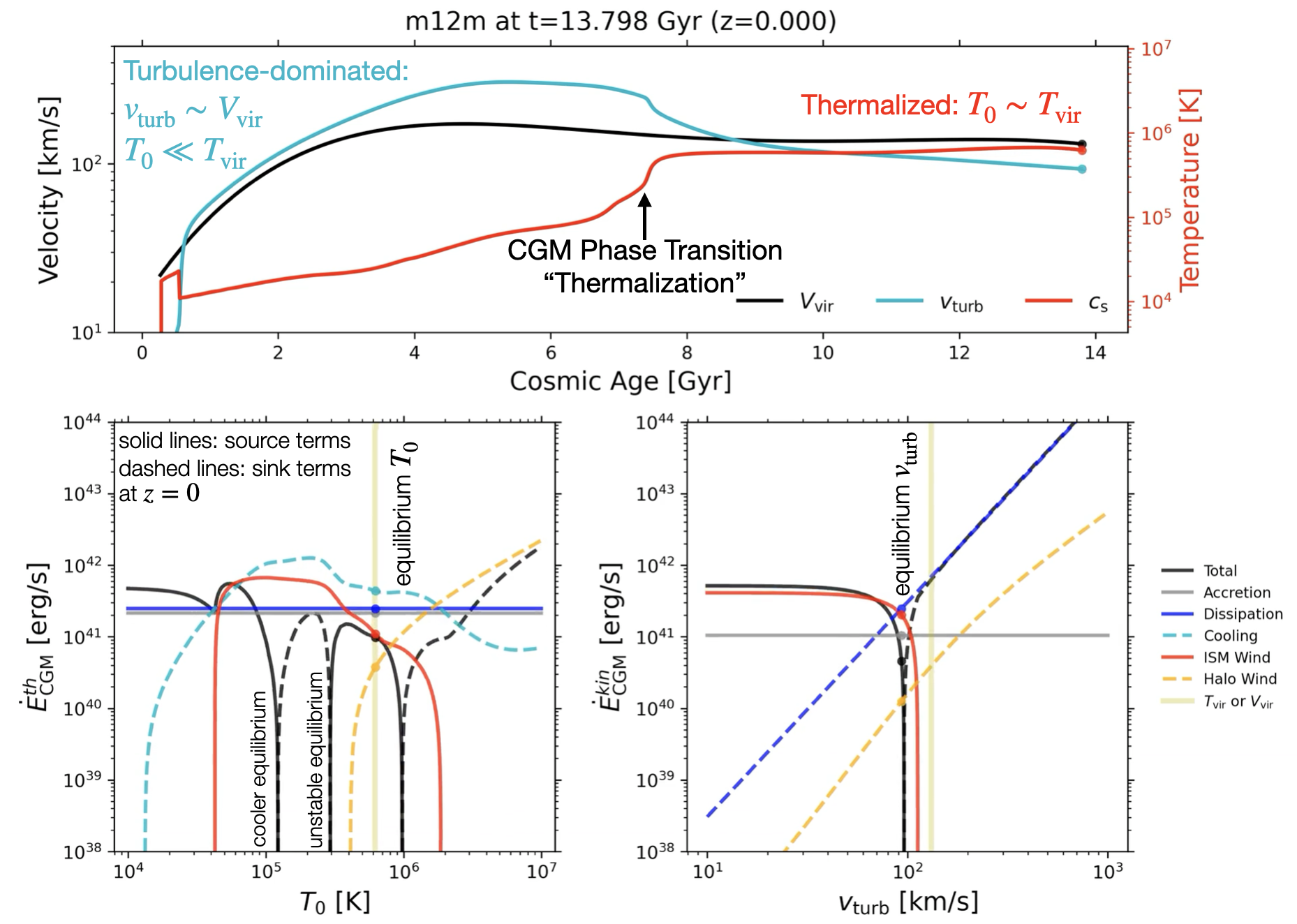}
\caption{A single annotated frame from our supplementary movie showing the time evolution of the equilibrium $T_0$ and $v_{\rm turb}$ for a MW-mass halo (m12m). The top panel shows the fiducial evolution of $c_{\rm s}$ (i.e., $T_0$) and $v_{\rm turb}$ relative to $V_{\rm vir}$. The bottom panels show how $\dot{E}_{\rm th}$ and $\dot{E}_{\rm kin}$ respectively depend on $T_0$ and $v_{\rm turb}$. We also plot the dependence of the individual $\dot{E}$ terms on $T_0$ and $v_{\rm turb}$ to help understand where the source terms (solid lines) balance the sink terms (dashed lines) causing the formation of equilibrium points. This frame is at $z=0$ where the main relevant equilibrium, which the model lies very close to, is at a slightly super-virial $T_0\approx10^6$ and slightly sub-virial $v_{\rm turb}\approx90$ km/s. The 50 second animated version of this figure illustrates the secular evolution of model equilibria due to forcing terms from cosmological accretion as well as bifurcations in the system as heating overcomes cooling at low temperatures during the thermalization process. We urge the reader to download our supplementary movie for this and other halos from the online version of the article.}
\label{fig:m12_frames}
\end{figure*}

The top panel of Figure \ref{fig:m12_frames} shows the time evolution of $v_{\rm turb}$ and $c_{\rm s}$ (or equivalently $T_0$) predicted by our model when we run it on the merger tree of a single MW-mass FIRE-2 halo (m12m). Also shown are the dependence of $\dot{E}_{\rm th}$ and $\dot{E}_{\rm kin}$ on varying $T_0$ and $v_{\rm turb}$, respectively, without any other parameter variations. We urge the reader to download our supplementary movie showing the time evolution of this figure which otherwise shows only a single annotated frame at $z=0$.

We see that at early times, the CGM temperature in our model is quite cool ($T_0\sim10^4-10^5$ K) but turbulent ($v_{\rm turb}\gtrsim V_{\rm vir}$). By $z=0$, the CGM temperature rises to $\gtrsim T_{\rm vir}$ and the turbulent velocity has decayed to $<V_{\rm vir}$. Thus this is another way to visualize the CGM phase transition that we described in the previous section, with the phase transition in this case happening abruptly at $t\approx7$ Gyr ($z\approx0.75$). At $z=0$ as shown in this frame, there is only one equilibrium for $T_0$ and $v_{\rm turb}$ and the model lives very close to it. The $T_0\approx10^6$ K equilibrium is set by a balance between the thermal energy sink terms (primarily $\dot{E}_{\rm cool}$ though $\dot{E}_{\rm out,halo}$ also plays a role) and the heating source terms (dissipation and accretion dominate over the wind thermal energy at $z=0$). Similarly, the $v_{\rm turb}\approx90$ km/s equilibrium is set primarily by a balance between turbulence dissipation (sink term) and turbulence driving by SN winds and cosmic accretion. 

\begin{figure}
\centering
\includegraphics[width=\hsize]{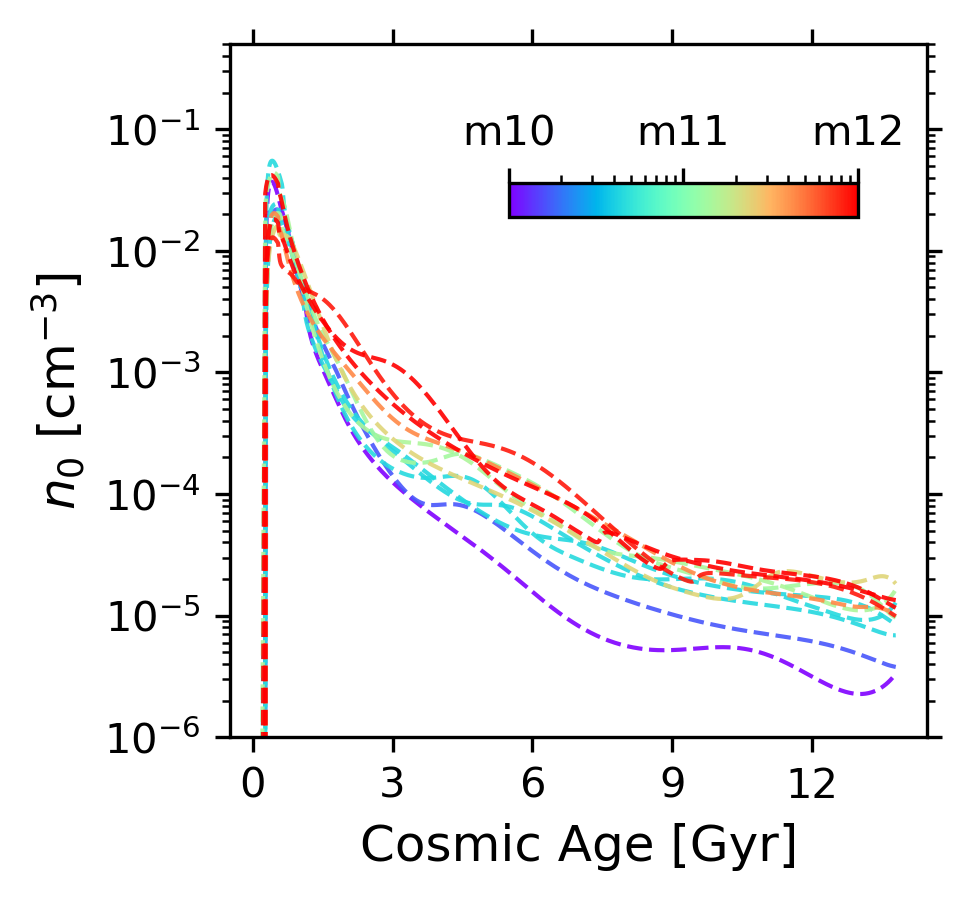}
\caption{The average CGM density ($n_0$ in Equation \ref{eqn:n0}) drops steadily with time reflecting the universal decrease in the matter density due to cosmic expansion. Since $\dot{E}_{\rm cool}\sim n^2\Lambda$ and the CGM mass and metallicity remain roughly constant with time (except for a steep early rise as halos first form), the radiative cooling rate of the CGM also steadily decreases with time as shown in the supplementary movies of Figure \ref{fig:m12_frames}. This gradual decrease in density determines when cooling at low temperatures can no longer keep up with heating thus driving up the equilibrium CGM temperature to $\sim T_{\rm vir}$.}
\label{fig:n0}
\end{figure}

In the movie version of Figure \ref{fig:m12_frames} (see caption for download link), we can compare the time evolution of the model to its equilibria. In general, the equilibria themselves evolve with time and the model generally follows this secularly evolving equilibrium. For example, at early times there is only a single equilibrium temperature and it steadily increases due to the forcing terms from cosmological accretion and the increasing star formation rate. However, just before and after the phase transition, we see bifurcations in the solution to our coupled ODE system: multiple equilibria can appear and disappear depending on the shape of the cooling function and where it intersects with the sum of the various heating terms. Just before the phase transition, the original cool equilibrium vanishes since the heating terms exceed the available cooling and the model is then forced to quickly evolve towards the next remaining hotter equilibrium. These bifurcations are largely driven by the fact that the $\dot{E}_{\rm cool}$ drops steadily with time up until the phase transition. Since $\dot{E}_{\rm cool}\propto n^2\Lambda$, this steady drop in cooling reflects the gradual decrease in the mean cosmic density as shown in Figure \ref{fig:n0}. We have verified that both $\Lambda$ and $Z_{\rm CGM}$ remain roughly constant with time and that it is indeed the decline in the CGM density that drives the reduction in $\dot{E}_{\rm cool}$. 

This general time-dependent picture also applies to our dwarf halos but the results are different (see our supplementary movies for dwarfs analogous to Figure \ref{fig:m12_frames}). In the intermediate-mass dwarfs, the equilibrium temperature remains close to the peak of the cooling curve until very late times so there is either no phase transition or it happens very late. In these classical dwarfs, the heating terms are not able to overcome cooling so we expect roughly $\sim10^5$ K temperatures. In contrast, low-mass dwarfs become thermal pressure-dominated at very early times in our model with their CGM temperature exceeding the halo virial temperature by a factor of $\sim2$. These ultrafaint-scale halos have low enough densities that their CGM is susceptible to photoionization heating from the UVB, which leads to a reduced net cooling rate that cannot balance heating from SN winds, turbulence dissipation and cosmic accretion except at a slightly super-virial temperature. In addition, we found in FIRE-2 and thus prescribed in our model that $f_{\rm thermal}^{\rm accretion}\approx0.8$ rather than $0.5$ for these $M_{\rm vir}\sim10^{10}M_{\odot}$ halos (see Figure \ref{fig:logistics}), and this contributes to them forming thermally-supported halos quite early.

\section{Discussion}

\subsection{Implications for the CGM--galaxy connection}\label{sec:implications}
We have presented a new model that is capable of predicting the time evolution of the global thermodynamic state of the CGM. This was accomplished by self-consistently linking both thermal and turbulent kinetic energy flows in the CGM to cosmic accretion and SN-driven galactic winds. In \citet{carr23}, we use the purely thermal limit of this model to understand the shape of the stellar-to-halo-mass (SMHM) relation, which has been empirically constrained by \citet[][see also the recent review by \citealt{wechsler18}]{behroozi19}. There, we found that the SMHM relation should be quite insensitive to variations in the mass loading factor of galactic winds alone because the CGM in our new framework is self-regulated. Increasing the mass loading factor while keeping the energy loading factor fixed leads to an increase in the CGM density and thus a decrease in the CGM specific energy, which in turn increases the mass accretion rate back into the ISM without achieving the intended decrease in SFR and hence SMHM ratio. In contrast, increasing the specific energy of the winds can lead to an increase in the CGM specific energy and therefore a suppression in the cooling rate and SFR. Although we do not perform a parameter space exploration and forward modeling of the SMHM relation in this paper with the additional turbulent CGM component, we expect very similar conclusions regarding the self-regulating nature of the CGM and implications for the SMHM relation. 

What is unique about our model is that it is single-phase but that phase changes naturally with time (we again urge the reader to watch our supplementary movies corresponding to Figure \ref{fig:m12_frames} to better understand how the individual $\dot{E}$ terms contribute to the evolution of model equilibria). The phase dichotomy is such that at early times, turbulence dominates the CGM energy budget because densities and thus radiative cooling rates are very high while dissipation rates are relatively low. Hence it is only natural that $E_{\rm CGM}^{\rm th}\propto T_0$ should remain relatively low at early times while $E_{\rm CGM}^{\rm kin}\sim v_{\rm turb}^2$ remains elevated for longer. This has the striking implication that the CGM--galaxy connection at early times should primarily be set by the turbulent rather than thermal properties of the CGM since the turbulence is the main source of pressure support and hence self-regulation for both the CGM and star formation. Of course, strong individual bursts of star formation can also cause jumps in the CGM thermal energy due to large-scale heating but the subsequent radiative cooling should be rather efficient, at least in the high-density conditions of the early Universe (whereas the turbulence may be more long-lived).\footnote{See our characterization of SN-driven winds in FIRE-2 by \citet{pandya21}, particularly the supplementary movies corresponding to their Figures 1 and 2.} 

We find that turbulence plays an important role at early times in the CGM of MW-mass halo progenitors as well as down to low redshifts for intermediate-mass (``classical'') dwarfs. However, ultrafaint dwarfs seem to have a much shorter lived early turbulent phase and instead transition to a warm thermal-pressure dominated CGM at quite high redshift. At first, this may appear natural because these low-mass dwarfs are assumed to have much smaller eddy sizes, but they also have proportionately smaller turbulent velocities so that the eddy turnover timescale (i.e., the ratio $R_{\rm turb}/v_{\rm turb}$) itself remains similar to that of more massive halos. Instead, it is simply that these halos lie below the peak of the cooling curve and their low CGM densities makes them susceptible to photoionization heating from the UV background, which together causes heating to exceed the available radiative cooling at an earlier time compared to the later phase transitions seen for more massive halos. The inefficient star formation histories and low SMHM ratios of ultrafaint progenitors thus appear to be intimately connected to the inability of their CGM to cool efficiently at early times. However, turbulence may still be important in limiting ISM accretion and excess star formation at very early times ($z\gtrsim3$) in ultrafaint progenitors when their CGM would otherwise have very short cooling times and be susceptible to thermal instabilities.

It is curious how closely connected the CGM phase transition is to the continuous decrease of the mean cosmic density $\rho_m$ for all halo masses we consider. We note that the CGM masses of our halos are relatively constant with time except for an initial steep rise as halos first build up. And yet the CGM number density $n(r)$ steadily declines with time, reflecting the importance of the growth of halo radii as the Universe expands. Since $\dot{E}_{\rm cool}\sim n^2\Lambda$, this means that the CGM cooling time gets progressively longer with time until it becomes the dominant term in limiting the ISM accretion rate and hence star formation and turbulence driving. This effect is even more pronounced in the ultrafaint progenitors for which the ratio of specific wind energy to specific CGM binding energy is preferentially larger. Thus they have reduced baryon fractions due to both preventative feedback suppressing halo gas accretion as well as their more easily overpressurized CGM ejecting a substantial fraction of previously accreted baryons. In turn the lower overall CGM densities of dwarfs contributes to their earlier CGM phase transition. With that said, density is not the only contributor to the phase transition: our assumption for $R_{\rm turb}$ directly sets the turbulence decay timescale so variations in this uncertain parameter also play an important role. 

While CGM turbulence plays a crucial role in regulating galaxy formation at early times, it is the physics of atomic cooling that governs the phase transition and CGM thermodynamics at late times. In particular, the temperature, density and metallicity will set the ionization state of different metals and in turn dictate the maximum achievable radiative cooling rate. When this cooling rate can no longer balance the heating from SN-driven winds, turbulence dissipation and cosmic accretion, we get a bifurcation in the solution to our ODEs in the sense that cool equilibria vanish and the system quickly evolves to the next hotter equilibrium state as imposed by the cooling curve. Even in the intermediate-mass dwarfs that retain significant CGM turbulent pressure support down to low redshift, it is only possible because they just happen to converge to a CGM (and virial) temperature that is close to the peak of the cooling curve ($\sim10^5$ K). This results in relatively low $t_{\rm cool,eff}/t_{\rm ff,eff}\sim1$ and places classical dwarfs squarely in the thermal instability regime, allowing star formation and turbulence driving to continue unimpeded.

Another way to appreciate our model is in the context of CGM virialization, though we argue ``thermalization'' is a more apt description since even at early times when $T_0\ll T_{\rm vir}$, we still have $v_{\rm turb}\sim V_{\rm vir}$ which satisfies the virial theorem (see Figure \ref{fig:m12_frames}). \citet{birnboim03} argued that gas accreting into halos more massive than $\sim10^{11}M_{\odot}$ gets shock-heated to the virial temperature of the halo, but that, in the absence of feedback, such shocks do not form in lower mass halos or at $z>2$ owing to efficient radiative cooling of the infalling gas. Subsequent studies using cosmological simulations demonstrated that the non-shocked gas accretes along dense, cold filaments and that this ``cold mode accretion'' directly feeds central galaxies and powers their star formation \citep{keres05,dekel09,keres09,fauchergiguere11,vandevoort11}. This picture is complicated by the contribution of galactic winds to the hot CGM \citep[e.g.,][]{vandevoort12,fielding17a}, the angular momentum and baryon fraction of halos \citep{stern20}, and potentially unresolved hydrodynamical instabilities that may cause cold gas filaments to disintegrate before they can reach the central galaxy \citep[e.g.,][]{mandelker20}. More recently, \citet{stern21} extended this standard CGM virialization (thermalization) picture to take into account radial dependence and directionality. They found that the CGM of MW-mass FIRE-2 halos virializes (thermalizes) gradually from the ``outside-in'' over a period of several Gyr, completing by $z\approx0.5$ (this is consistent with our own measurements of $f_{\rm thermal}^{\rm CGM}$ from the particle data in Figure \ref{fig:energetics}). This ``inner CGM virialization'' (thermalization) is correlated with the formation of a thin disk, the transition from bursty to steady star formation and damping of the strength of SN-driven winds \citep[see also][]{anglesalcazar17b,pandya21,gurvich22,hafen22}. Thus the conditions of the inner CGM are important in setting the properties of the central galaxy. In contrast, intermediate-mass FIRE-2 dwarfs retain a thermally unstable inner CGM down to $z\sim0$ (even though their outer CGM is thermalized) and do not show signs of a prominent stable gaseous disk \citep[compare Figures 4 and 5 in][]{stern21}. 

Although our model treats the entire CGM as a single zone and is therefore not designed to address the radial dependence of CGM thermalization (a radially continuous 1D model would be needed for that) and consequently predicts a more abrupt phase transition than measured in FIRE-2, we are consistent with the overall picture of CGM thermalization. For our fiducial choice of model parameters calibrated from the FIRE-2 simulations, MW-mass halos transition from an early, turbulent, thermally unstable CGM to a warm/hot thermally-supported CGM with temperature only slightly above $T_{\rm vir}$ by $z\approx0.5$ as in FIRE-2 (although we note that these results would change for different parameter choices). The CGM of intermediate-mass dwarfs does not experience this phase transition until very late times, if at all. However, we predict that the CGM of ultrafaint dwarfs should experience the phase transition at much higher redshift contrary to the simple statement that any halo with $M_{\rm vir}\lesssim10^{11}M_{\odot}$ should always experience cold mode accretion. The low baryon fractions, low metallicities and reduced dense ISM gas fractions of these dwarfs invites additional complications such as significant CGM photoionization heating and easier SN wind breakout which our model accounts for phenomenologically with our parameterized loading factors and UV background-dependent cooling function. 

In the future, we will use our new framework to explore whether the observable properties of galaxies themselves are correlated with the observable properties of their CGM. This will require forward modeling predictions for large populations and assessing the scatter in galaxy properties at fixed CGM properties and vice versa. We expect the CGM--galaxy connection to manifest in several observable scaling relations such as the mass--metallicity relations for both stars and ISM gas \citep[and possibly also the CGM although that may be harder to untangle observationally, but see][]{faerman22}, the stellar-to-halo mass relation and ISM gas fractions \citep[see our initial exploration of these in][]{carr23}, and the star-forming main sequence and the cause of its scatter \citep[e.g.,][]{rodriguezpuebla16}. It will be very informative to see how sensitive our predictions are for these and other galaxy--CGM scaling relations as we vary our free parameters, especially for model realizations far from our fiducial FIRE-calibrated one. We also plan to investigate whether adding CGM-related observational constraints to the usual set of galaxy population metrics at $z=0$ breaks parameter degeneracies and shrinks the allowed parameter space of our model.

\subsection{Comparison to previous physical models} 
As reviewed by \citet{carr23}, our model builds on but also goes significantly beyond previous ``bathtub'' and semi-analytic modeling approaches. Bathtub models generally only deal with the ODEs for the evolution of ISM mass, stellar mass and the metallicities of these two components, and a special subset of bathtub approaches further enforce an equilibrium condition on the ISM mass (i.e., set $\dot{M}_{\rm ISM}=0$ assuming inflows balance star formation and outflows). These models generally neglect the physics of the CGM and instead have free parameters that roughly control how much cosmic accretion ends up as cold gas in the galaxy and becomes available for star formation. While this approach has achieved success in reproducing some key observed scaling relations of galaxies at a range of redshifts \citep[see the recent review by][]{tacconi20} and even summarizing the gas flow cycle in hydrodynamical simulations \citep[e.g.,][]{neistein12,mitchell22}, it leaves much to be desired in terms of clearly elucidating the detailed underlying physics. In contrast, we are explicitly accounting for multiple physical processes that regulate the CGM not only via an ODE for its mass evolution but also its thermal and turbulent kinetic energy evolution. The closest model to ours is the purely thermal limit version that we have put forward in \citet{carr23} which we would converge to as $f_{\rm thermal}^{\rm wind}\to1$ and $f_{\rm thermal}^{\rm accretion}\to1$ \citep[see also][which similarly tracks the thermal energy of the hot CGM phase]{cousin15}. One difference is in how we set our model parameters: \citet{carr23} use a fixed power law for the mass loading factor, adopt an observationally-inferred relation for the ISM depletion time, and fit for the parameters of a power law for the energy loading factor with halo mass using the \citet{behroozi19} SMHM relation as a constraint. In contrast, we constrain these and other free parameters directly from the FIRE-2 simulations as advocated in \citet{pandya20} and \citet{pandya21}. In the future we plan to do an exhaustive parameter space exploration and comparison to observations in the spirit of \citet{carr23} but that will require sophisticated inference machinery that is beyond the scope of this paper. 

SAMs include many more physical processes than both bathtub models and our new approach such as satellite orbital dynamics and gas stripping, merger-induced starbursts, recycling of halo outflows back into the CGM, growth of and feedback from supermassive black holes, multiphase ISM partitioning and galaxy structural evolution \citep[see][for a recent review]{somervilledave15}. It is possible that including some of these additional processes could affect the thermodynamic evolution of the CGM within our new framework. For example, the extra energy injected into the CGM from winds and turbulence stirred by satellites and black holes could affect the nature of our CGM phase transition, especially in more massive halos. And while the standard SAM treatment of many of these extra physical processes may itself be subject to uncertainties, we expect our new CGM model to integrate naturally within the foundation of existing SAMs. Furthermore, \citet{pandya20} showed that existing CGM prescriptions in some SAMs can predict dramatically different CGM properties compared to simulations like FIRE-2 (e.g., orders of magnitude lower CGM masses for dwarfs in their Figure 6). Even though our new model shows its own discrepancies relative to FIRE-2, it is an improvement over previous approaches and it follows the overall trends of the simulations. We therefore intend for the new model presented in this work to become the backbone of a next-generation SAM that we ourselves are building to which many of the aforementioned uncertain physical processes will be added in piece by piece.

We point out that there have been many previous efforts to revamp the way that SAMs traditionally model the CGM. As nicely reviewed by \citet{lu11}, most existing implementations of the CGM in SAMs trace back to \citet{whitefrenk91} who assumed that the thermodynamics of the CGM traces that of the underlying dark matter in the sense that the CGM temperature everywhere must be equal to the halo virial temperature. A singular isothermal density profile ($n\propto r^{-2}$) for the CGM is typically assumed and the metallicity is also assumed to be the same everywhere, usually with solar abundance ratios. For simplicity and historical reasons the \citet{sd93} cooling function is commonly assumed to compute the radiative cooling rate, but this assumes CIE even though photoionization heating by the UV background can be important in halo outskirts and especially around dwarfs \citep[see also][]{benson02}. The turbulent kinetic component is altogether generally neglected. These simplifying assumptions allow one to compute a so-called ``cooling radius'' within which the cooling time of the gas is shorter than some long timescale such as the Hubble time. That cooled gas is then assumed to free-fall into the ISM on the halo dynamical timescale. Whenever $R_{\rm cool}>R_{\rm vir}$, it is assumed that the CGM is undergoing ``cold mode'' accretion and that filaments are directly free-falling into the ISM from large scales \citep[e.g.,][]{guo11}. There are further extensions built on top of this approach such as tracking CGM angular momentum evolution \citep[e.g.,][]{stevens17,hou18,lagos18} and allowing for the simultaneous cosmic accretion of cold and hot gas \citep{lu11,benson11,cousin15}. 

What \citet{carr23} and we have done is to show how the thermodynamic evolution of the CGM can be decoupled from that of the dark matter: by introducing the $\dot{E}_{\rm CGM}^{\rm th}$ and $\dot{E}_{\rm CGM}^{\rm kin}$ ODEs as well as free parameters for the density, temperature and turbulent velocity structure of the CGM, we can track energy flows in the CGM and predict its global thermodynamic state \citep[see also][for another approach in the purely thermal limit]{cousin15}. We roughly converge to the assumption that the temperature of the CGM is $\approx T_{\rm vir}$ at late times in MW-mass halos, but at early times when cooling rates are very high, we generally predict sub-virial temperatures that are then self-consistently fed into the \citet{wiersma09} cooling function which accounts for CGM photoionization. During these early times when CGM cooling times are extremely short, we also suggest that turbulent pressure can significantly limit ISM accretion and hence prevent excess early star formation. SN-driven winds are the dominant source of CGM turbulence except at very early times before SF kicks in, when cosmic accretion would be the sole driver (see bottom-right panel of the movie version of Figure \ref{fig:m12_frames}). This may have implications for the normalization and slope of the faint end of the stellar mass function and mass--metallicity relation for both stars and the ISM. In addition, whereas SAMs generally assume a phenomenological function to predict a mass outflow rate from the halo depending on the virial velocity of the halo, our formalism provides a prediction of the excess energy and mass that must be vented by an overpressurized CGM to remain in a quasi-hydrostatic equilibrium. As suggested by the models of \citet{lu15}, \citet{pandya20} and \citet{carr23}, this outflowing energy from the halo may act in concert with other mechanisms to pre-heat the gas outside of low-mass halos and prevent its accretion, further contributing to the reduced baryon fractions and SMHM ratios of dwarfs.  

Finally, we briefly remark on the existing family of 1D CGM models that are very compelling in their ability to describe the properties of the CGM in both simulations and observations. These models envision three different physical scenarios for the CGM: steady-state cooling flows \citep[see][and references therein]{fabian94,stern19}, hydrostatic equilibrium \citep[e.g.,][]{faerman17,qu18,faerman20}, and precipitation \citep[e.g.,][]{mccourt12,sharma12,voit15}. The advantage of these models is that they can predict CGM observables starting with very clear explanations for the underlying physical principles. Their disadvantage is that these models do not simultaneously model the galaxy formation process and are generally only applied at a single instant (though their parameters can be varied to describe CGM conditions in different mass halos or at a range of redshifts). Our approach is complementary in that we explicitly take into account time dependence via our system of coupled ODEs that self-consistently links energy flows in the CGM to cosmic accretion and SN-driven winds. However, one limitation of our approach is that we currently prescribe rather than predict the radial structure of the CGM: the slopes of the density and temperature profiles in our model are free parameters whereas HSE, precipitation and cooling flow models make clear predictions for these radial gradients. Another limitation is that our model is single-phase and does not account for scatter in the thermodynamic properties of the CGM as a function of radius whereas precipitation models at least attempt to capture the multiphase aspect of halo gas \citep[see also][]{esmerian21}. Extending our model to include a multiphase CGM is non-trivial but will be the subject of future work.

In the future it would be interesting to forward model observables of the CGM such as column densities of various ions as a function of impact parameter, X-ray luminosities and the Sunyaev-Zel'dovich effect. We could do this with our fiducial assumed density and temperature profiles but we could also feed the scatter in CGM properties predicted by our model at fixed halo mass into the HSE, precipitation and cooling flow models to see what those frameworks would predict for the CGM structure and related observables. For an initial exploration of this approach, we refer the reader to \citet{faerman22} who took the CGM masses and metallicities for a large population of MW-mass halos from the Santa Cruz SAM \citep{somerville15}, generated several CGM observables by exploring the parameter space of the \citet{faerman20} HSE model, and placed constraints on those HSE parameters by comparing to observations of the MW CGM \citep[see also][]{qu18}.

\subsection{Limitations and uncertainties}\label{sec:limitations}
Although we believe our new approach to be significant step towards a more self-consistent and predictive SAM of galaxy formation, our model is still subject to many limitations and uncertainties. Here we briefly provide a non-exhaustive list of possible issues and hence avenues for future work (grouped into a few representative categories): 

\begin{enumerate}
\item Turbulence: the main uncertainty is that we do not know how $R_{\rm turb}$ should vary with halo mass, redshift and CGM/galaxy conditions. In addition, our calculation of the turbulence dissipation rate is based on the largest eddy turnover time (Equation \ref{eqn:tffeff}) which is appropriate for subsonic turbulence but not necessarily for supersonic turbulence in which shocks may allow the turbulence to dissipate even faster \citep[but see][]{maclow99}. Relatedly, we do not distinguish between bulk flows and turbulence because we argue that any differences are effectively averaged over in our smooth model and because our model does predict bulk outflows of kinetic energy from the halo when $v_{\rm turb}>V_{\rm vir}$. However it would be good to check this with a stochastic model for bulk flows and explicitly introduce free parameters to separate the two processes. Finally, it would be insightful to develop predictive models for $f_{\rm thermal}^{\rm wind}$ and $f_{\rm thermal}^{\rm accretion}$ since these parameters largely control the amount of turbulence driving. We note that previous studies like \citet{birnboim03} predict $f_{\rm thermal}^{\rm accretion}$ based on accretion shock calculations but they neglect feedback which can dramatically alter outer CGM properties especially around dwarfs.
\item Atomic cooling physics: we are assuming the cooling function of \citet{wiersma09} which is a significant improvement over the \citet{sd93} tables generally adopted by SAMs because it allows us to take into account CGM photoionization heating. However, it would be good to check how our radiative cooling rates would be affected if we use more recent cooling tables \citep[e.g.,][which uses the more recent UV background model by \citealt{fauchergiguere20}]{ploeckinger20}. In the same vein, there may be updated parameterizations for the fraction of gas around dwarfs that is photoionized by the UV background and hence prevented from accreting \citep[compared to our assumed values from][which uses the old \citealt{haardt01} UV background model]{okamoto08}. Finally, non-equilibrium ionization processes are not captured by our assumed cooling function but may have significant effects on our predictions \citep{tumlinson17}
\item Multiphase gas: our CGM model is single-phase even though we expect gas to exist at a range of temperatures. It may be interesting to separately track the already-cooled gas and explore various models for its evolution \citep[e.g., cold cloud scenarios;][]{maller04,faerman23}. We also do not currently allow for the possibility that some fraction of cosmic accretion enters the CGM as filaments and directly deposits cold gas into the ISM without being subject to our normal energy flow cycle, which may be especially important at high-redshift \citep[e.g.,][]{mandelker20}. The inclusion of multiphase gas may also affect the nature of the phase transition including if, when and how abruptly it happens.
\item Radial structure: we only track the global thermodynamic state of the CGM but it is likely that the inner and outer CGM have quite different properties. Indeed \citet{stern21} find that the thermalization of the CGM in FIRE-2 has a radial dependence and directionality. Including even a steady-state (non-time-varying) radially-resolved 1D CGM model \citep[e.g.,][]{stern19} within our framework could help capture a more gradual CGM thermalization process and also account for changes in the gravitational potential energy as material moves in the CGM. A radially-resolved model would also negate the need for introducing a single arbitrary radius at which to define the effective free-fall time and instead allow us to compute $\dot{M}_{\rm cool}$ as a radial integral instead of Equation \ref{eqn:mdot_cool}.
\item Chemical evolution: our instantaneous recycling assumption is successful in roughly reproducing the CGM, ISM and stellar metallicities of the FIRE-2 halos as a function of time. We assumed that winds have the same metallicity as the ISM but this is likely not the case in reality (or even in detailed simulations like FIRE). We also parameterized the halo inflow metallicity from FIRE-2 but this could be modelled more self-consistently with an outer halo wind recycling model. More generally, we should switch to time-dependent, multi-yield functions to track the production and inflows/outflows of individual elements which would allow us to self-consistently predict abundance ratios of the gas and stars and assess the impact of assuming solar abundance ratios on CGM radiative cooling rates. 
\item Stochastic effects: our star formation model is continuous since we simply define a single ISM depletion time parameter, but star formation is a stochastic process, especially in dwarfs, in FIRE-2 and other high-resolution simulations \citep[e.g.,][]{muratov15,christensen16,sparre17,fauchergiguere18,iyer20,pandya20,gurvich22}. Thus we should account for stochastic bursts of feedback and hence heating effects on the CGM. In particular, winds and halo outflows should occur on a variety of timescales. Related to this is the effect of recycling: some fraction of winds may recycle in the inner halo on a rapid timescale as fountain flows and some fraction of halo outflows may also recycle back into the CGM \citep{anglesalcazar17}.
\item Other missing physics: connection to galaxy structure and specifically disk formation \citep{forbes19},  multi-phase gas and turbulence in the ISM \citep{ginzburg22,forbes23}, a self-consistent star formation model, cosmic rays and magnetic fields as additional sources of non-thermal pressure support beyond turbulence alone, additional energy input from supermassive black hole feedback which we expect to be important for extending this model to group/cluster scales, wind heating, dynamical friction heating, turbulence driving, stochastic feedback from major mergers and contribution of multiphase CGM gas by satellites.
\item Measurement uncertainties: in any analysis of simulations or observations, there will be uncertainties in derived quantities due to the definitions and techniques that are adopted. Our measurements of model parameters and galaxy/CGM properties in the FIRE-2 simulations rely in some cases on arbitrary choices for our chosen definition of where the CGM begins and ends ($0.1-1.0R_{\rm vir}$), our neglect of satellite contributions to host CGM heating and cold gas content, our choice of spherical boundaries through which to track mass, energy and metal fluxes into and out of galaxies/halos ($0.1R_{\rm vir}$ and $R_{\rm vir}$), our decision to not impose any further velocity cuts except $v_{\rm rad}=0$ as the split between inflowing and outflowing particles, etc. Also, while our approach of adding up the thermal energy of all CGM particles should be robust, our measurement of the total CGM kinetic energy is an upper limit to the turbulent CGM energy component alone since bulk flows and rotation can be important. We also have not yet performed multi-snapshot particle tracking to understand the longer term evolution of particles such as whether inflows actually get to the galaxy and what fraction of halo outflows recycle back into the CGM versus become unbound from the halo forever \citep[but see][]{hafen20}.
\item Statistical inference: we have not exhaustively explored our parameter space but instead have fixed most of our parameters to the values we measure in the FIRE-2 simulations. However, many of these parameters likely suffer from degeneracies and uncertainties, and probably vary across suites of simulations with different implementations of key physical processes. It would be good to measure the parameters of our model in many different simulations and assess the scatter as a measure of uncertainty on galaxy formation-related processes. Longer term, it would be informative to use sophisticated techniques such as implicit likelihood inference \citep{2020PNAS..11730055C} to explore the highly multi-dimensional parameter space of our model using both simulations and observations as constraints. In particular, it would be good to know what extra value CGM observations bring for breaking parameter degeneracies in our model and SAMs more generally compared to just using galaxy scaling relations alone. 
\end{enumerate}

\section{Summary}
We have presented a new time-dependent two-zone model for the co-evolution of galaxies and their CGM. Our model self-consistently tracks the evolution of the global thermal energy and turbulent kinetic energy of the CGM accounting for energy input from SN-driven winds and cosmic accretion as well as radiative cooling, turbulence dissipation, and large-scale halo outflows when the CGM becomes overpressurized. We explore the dynamics of the model with a particular focus on the phase transitions that occur and the processes that drive them. In a companion paper by \citet{carr23}, we use the purely thermal limit of this kind of model to instead understand the shape of the stellar-to-halo-mass relation and ISM gas fractions from an empirical perspective, thus demonstrating that our approach can be a powerful way to connect to both more sophisticated theoretical models as well as observations. Our main takeaways are as follows:

\begin{itemize}
\item By self-consistently tracking both the thermal and turbulent kinetic energy flows in the CGM, we can decouple the thermodynamics of the CGM from that of simple virial arguments. In particular, the average global temperature of the CGM need not always be equal to the virial temperature of the halo, and short CGM cooling times need not lead to direct freefall of gas into the ISM since it can still be supported by turbulent pressure.
\item The model predicts that the CGM can undergo a phase transition (``thermalization'') from an early, cool turbulent phase to a warm/hot roughly virial temperature volume-filling phase at later times. The phase transition in our model appears closely related to the ever-decreasing mean density of the universe and hence CGM. Since $\dot{E}_{\rm cool}\propto n^2 \Lambda$, eventually cooling drops sufficiently low that it cannot keep up with the various heating terms, which causes the cool equilibrium to vanish and forces the system to quickly evolve towards the remaining hotter equilibrium at $\sim T_{\rm vir}$ (this is a bifurcation of the ODE system). As SFRs and wind specific energies decrease with time, there is less driving of turbulence in the CGM so it generally decays. The assumed size of the largest turbulent eddies in the CGM is also an important ingredient for the CGM thermalization process.
\item The equilibrium solutions to our system of ODEs (which the model will tend to evolve towards in the absence of forcing terms from cosmological accretion) are sensitive to model parameters such as the specific energy of galactic winds and the size of the largest turbulent eddies in the CGM (which determines the turbulence dissipation rate). Simple parameter space exploration for idealized Milky Way parameters at $z=0$ shows that increasing the specific energy of galactic winds would lead to a higher equilibrium thermal temperature and turbulent velocity as more energy is pumped into the CGM (and vice versa). On the other hand, decreasing the largest turbulent eddy size leads to a decrease in the equilibrium turbulent velocity and an increase in the CGM temperature as the turbulence decays and dissipates as heat more quickly. A more exhaustive parameter space exploration is deferred to future work.
\end{itemize}

We then performed an initial calibration of the model by measuring many of its free parameters from the FIRE-2 cosmological hydrodynamical ``zoom-in'' simulations \citep{hopkins18}, namely the ISM depletion time, wind mass loading factor, wind specific energy, halo inflow metallicity, halo gas accretion efficiency and thermalization of accretion and wind energy. For model parameters that could not be directly constrained from the simulations such as the turbulence dissipation timescale and gas infall timescale, we make reasonable fiducial assumptions. In particular, we assume that the largest CGM eddy size is of order the halo virial radius at early times but drops to the inner halo radius at late times (based on the argument that the primary driver of turbulence transitions from cosmic accretion to SN-driven winds). This last assumption needs to be checked with future analysis of the simulations. We find that:

\begin{itemize}
\item The model approximately captures the general trends in the simulations in terms of the mass assembly histories of the CGM, ISM and stars for a wide range of halos from ultrafaints to MW-mass halos. The main discrepancies are that our CGM masses tend to be a factor of $\sim2$ lower than in the simulations, and some dwarfs show up to $\sim10$ times higher stellar and ISM masses in our model than in the simulations. We argued that accounting for halo-to-halo scatter in our measurements of model parameters from the simulations and perhaps varying our parameterizations for physical processes that were not directly constrained from the simulations (e.g., turbulence dissipation and pressure support) may help alleviate these differences.
\item The model reproduces the overall halo baryon fractions in the simulations and explains why dwarfs only have values of $\sim10-50\%$: dwarfs in the model do not accrete their full complement of the cosmic baryon fraction and also eject a significant fraction of previously accreted baryons via outflows from their overpressurized CGM. We showed that the CGM overpressurization channel becomes increasingly important towards lower mass halos.
\item The bulk metallicities of the CGM, ISM and stars are also generally in agreement except that the MW-mass halos at $z\sim0$ have higher CGM metallicities in our model than FIRE-2 and the lowest mass dwarfs at high-redshift have higher metallicities for all three components in our model compared to the simulations. We suggested that varying the wind enrichment factor and using time-dependent rather than instantaneous recycling can improve these discrepancies.
\item In addition to mass budgets, the model predicts inflow and outflow rates of gas mass, metals and energy that are roughly in agreement with measurements from the simulations. The main disagreement in this context occurs for the lowest mass dwarfs for which our model tends to predict higher flow rates. We argued that changing our uncertain parameters related to the gas infall and turbulence dissipation timescales (not directly constrained by the simulations) can improve the discrepancy for the dwarfs.
\item With the FIRE-2 parameters, the model predicts that the CGM phase transition happens at high-redshift for ultrafaint dwarfs and at low redshift (if at all) for intermediate-mass dwarfs. This global thermalization of the CGM is clearest in the MW-mass halos where, for our chosen parameters, it happens at $z\approx0.5$. We find that this CGM thermalization also occurs in the FIRE-2 simulations but that it is more gradual, likely due to it being a radially dependent process \citep[as shown by][]{stern21}. We argue that the phase transition is more abrupt in our model because it treats the entire CGM as a single zone.
\end{itemize}

While our model is expressive enough to roughly reproduce the simulations, the discrepancies above point to its limitations. We discussed several physical processes that are currently neglected but argued that the model is ripe for future extensions. This includes multi-zone CGM modeling, inclusion of multi-phase gas, additional sources of non-thermal pressure support in the CGM beyond turbulence such as cosmic rays, and other aspects clearly listed in section \ref{sec:limitations}.

\begin{acknowledgements}
We thank Lucy Reading-Ikkanda at the Simons Foundation for creating the illustration of our model in Figure \ref{fig:structure}, the FIRE team for kindly providing their simulation data, and the Scientific Computing Core at the Flatiron Institute for maintaining the supercomputer on which much of this work was performed. We also greatly thank Mark Voit, Fruzsina Agocs, Zirui Chen and Yakov Faerman for insightful discussions as well as the anonymous referee for helpful suggestions. Support for VP was provided by NASA through the NASA Hubble Fellowship grant HST-HF2-51489 awarded by the Space Telescope Science Institute, which is operated by the Association of Universities for Research in Astronomy, Inc., for NASA, under contract NAS5-26555. GLB acknowledges support from the NSF (AST-2108470, XSEDE grant MCA06N030), NASA TCAN award 80NSSC21K1053, and the Simons Foundation (through grant 822237 and support of the Learning the Universe collaboration). CAFG was supported by NSF through grants AST-1715216, AST-2108230, and CAREER award AST-1652522; by NASA through grants 17-ATP17-0067 and 21-ATP21-0036; by STScI through grants HST-AR-16124.001-A and HST-GO-16730.016-A; by CXO through grant TM2-23005X; and by the Research Corporation for Science Advancement through a Cottrell Scholar Award. DAA acknowledges support by NSF grants AST-2009687 and AST-2108944, CXO grant TM2-23006X, and Simons Foundation award CCA-1018464. JS was supported by the Israel Science Foundation (grant No. 2584/21).
\end{acknowledgements}

\bibliographystyle{aasjournal}
\bibliography{references}

\appendix 

\section{The purely thermal limit of our model}\label{sec:appendix}
\begin{figure*}
\centering
\includegraphics[width=0.9\hsize]{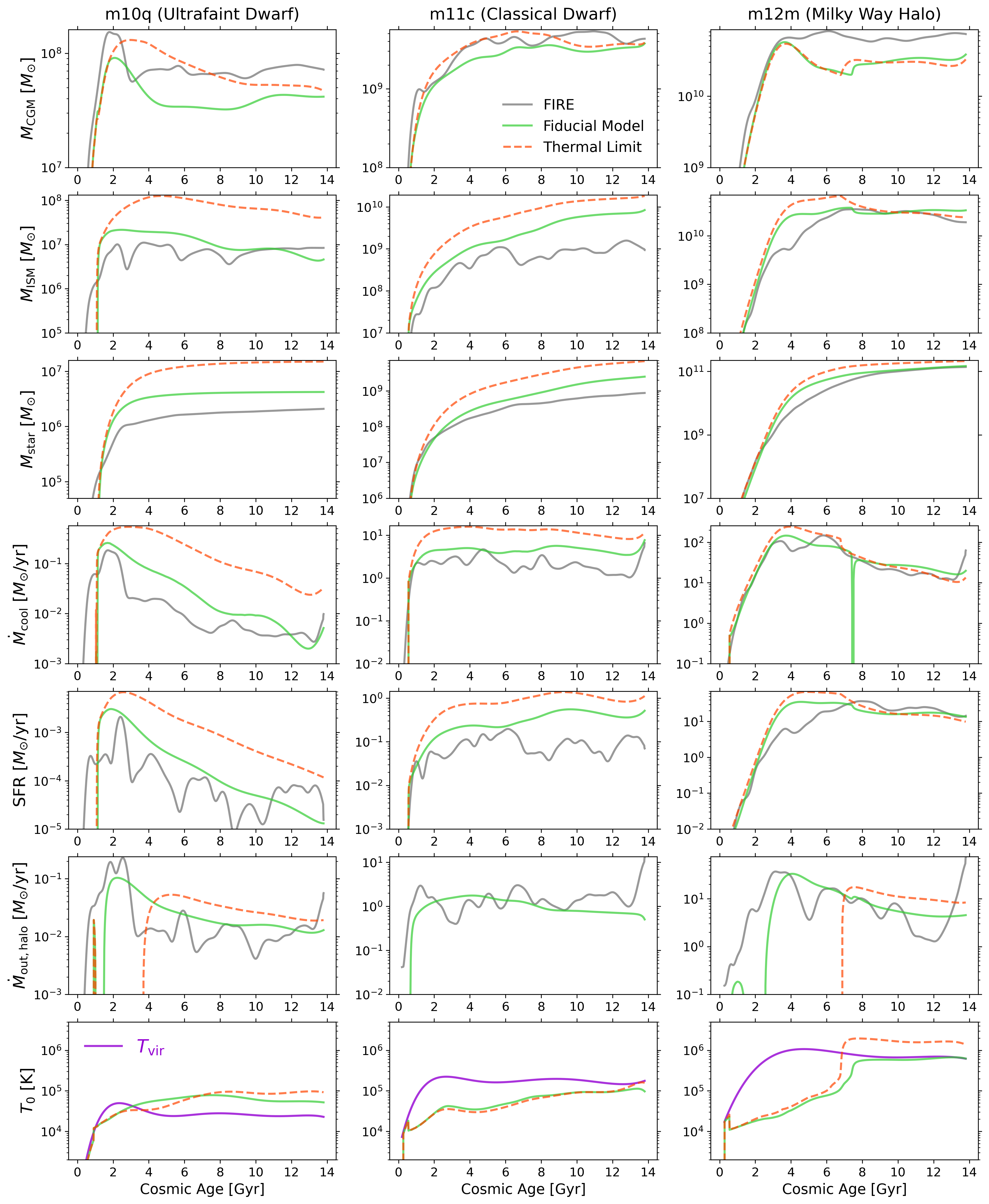}
\caption{Comparing our fiducial model to its purely thermal limit (i.e., no turbulence). Each column shows a representative halo from a different mass bin (left to right: ultra-faint dwarf, classical dwarf, MW-mass halo). Each row shows the time series of a key property (top to bottom: CGM, ISM and stellar mass, ISM accretion rate, SFR, halo mass outflow rate, average CGM temperature). Solid gray lines show our measurements from FIRE-2, solid green lines show predictions from our fiducial turbulent model, and dashed red lines show our thermal limit predictions. The purely thermal realization leads to higher ISM accretion rates, SFRs, and ISM and stellar masses as well as an under-pressurized CGM without halo outflows at early times.}
\label{fig:appendix}
\end{figure*}

If we set $f_{\rm thermal}^{\rm accretion}=1$ and $f_{\rm thermal}^{\rm wind}=1$, we are in the purely thermal limit of our model as in our companion paper by \citet{carr23}. In this case, there is no driving of turbulence by either cosmic accretion or SN winds and the CGM is supported only by thermal pressure. Keeping all other parameters fixed, Figure \ref{fig:appendix} shows how the time series of several key properties are affected in the purely thermal limit for one representative ultrafaint dwarf, classical dwarf and MW-mass halo. Neglecting turbulence generally leads to higher ISM accretion rates, SFRs and thus higher ISM and stellar masses (red dashed lines) compared to our fiducial turbulent model (solid green line) and FIRE-2 (solid gray lines). The differences are especially pronounced for the dwarfs. The lack of turbulence at early times also means that the CGM cannot be overpressurized since the high densities and cooling rates still lead to $T_0\ll T_{\rm vir}$. Thus there are no halo outflows at early times, and in the case of the classical dwarf which has maximally efficient cooling at all times, the CGM is never overpressurized and there are no halo outflows at all. The MW-mass halo still shows a phase transition but the post-thermalization temperature is consistently super-virial with $T_0\sim 2T_{\rm vir}$.

This exercise strongly suggests that turbulence is a necessary component of our model. Without turbulence, it seems difficult to accommodate the evolutionary histories of the FIRE-2 galaxies assuming our fiducial calibration of the other parameters. On the other hand, a more exhaustive exploration of the remaining parameter space is warranted since there will be degeneracies and since we showed in \citet{carr23} that the purely thermal model is capable of reproducing observed ISM gas fractions and the stellar-to-halo-mass relation with sufficiently high specific energy SN winds. We defer additional parameter variations and extensions of the model to future work.

\end{document}